\documentclass[9pt,shortpaper,twoside,web]{ieeecolor}

\usepackage{generic}
\usepackage{cite}
\usepackage{graphicx}
\usepackage{textcomp}
\usepackage{epstopdf}

\usepackage{array}
\newcolumntype{P}[1]{>{\centering\arraybackslash}p{#1}}
\usepackage{float}
\usepackage{tabu}

\usepackage{graphicx}
\usepackage{blindtext}
\usepackage{graphics} 
\usepackage{epsfig} 

\usepackage{amsmath}
\usepackage{amssymb}  
\usepackage{amsfonts}
\usepackage{mathrsfs}
\usepackage{mathptmx} 
\usepackage{mathtools}

\usepackage{times} 

\usepackage{multirow}
\usepackage{subcaption}
\usepackage{csquotes}

\usepackage{algorithm}
\usepackage{algpseudocode}
\usepackage[algo2e]{algorithm2e}

\usepackage{tikz}
\usetikzlibrary{shapes.geometric, arrows}
\usepackage{longtable}

\usepackage{amsbsy}
\usepackage{hyperref}
\usepackage{authblk}
\usepackage{leftidx}
\usepackage{ragged2e}
\usepackage[font={scriptsize}]{caption}
\usepackage{color}

\newtheorem{thm}{Theorem}
\newtheorem{lem}{Lemma}
\newtheorem{prop}{Proposition}
\newtheorem{cor}{Corollary}

\def\BibTeX{{\rm B\kern-.05em{\sc i\kern-.025em b}\kern-.08em
    T\kern-.1667em\lower.7ex\hbox{E}\kern-.125emX}}
\begin{document}

\title{A Convex Programming Approach to Data-Driven Risk-Averse Reinforcement Learning }

\markboth{ }
{Han \MakeLowercase{\textit{et al.}}:A convex Programming Approach to Data-Driven Risk-Averse Reinforcement Learning}
\author{Yuzhen Han, Majid Mazouchi, Subramanya Nageshrao and Hamidreza Modares 
\thanks{
Y Han, M Mazouchi and H Modares are with  the Department
of Mechanical Engineering, Michigan State University, East Lansing, MI, 48863, USA, (e-mails: hanyuzh1@msu.edu;  mazouchi@msu.edu; modaresh@msu.edu). S. Nageshrao is with Ford Research and
Innovation Center, Ford Motor Company, Palo Alto, CA 94304, USA,
(e-mail: snageshr@ford.com).}
}

\maketitle

\begin{abstract}
This paper presents a model-free reinforcement learning (RL) algorithm to solve the risk-averse optimal control (RAOC) problem for discrete-time nonlinear systems. While successful RL algorithms have been presented to learn optimal control solutions under epistemic uncertainties (i.e., lack of knowledge of system dynamics), they do so by optimizing the expected utility of outcomes, which ignores the variance of cost under aleatory uncertainties (i.e., randomness). Performance-critical systems, however, must not only optimize the expected performance, but also reduce its variance to avoid performance fluctuation during RL's course of operation. To solve the RAOC problem, this paper presents the following three variants of RL algorithms and analyze their advantages and preferences for different situations/systems:  1) a one-shot static convex program -based RL, 2) an iterative value iteration (VI) algorithm that solves a linear programming (LP) optimization at each iteration, and 3) an iterative policy iteration (PI) algorithm that solves a convex optimization at each iteration and guarantees the stability of the consecutive control policies. Convergence of the exact optimization problems, which are infinite dimensional in all three cases, to the optimal risk-averse value function is shown. 
To turn these optimization problems into standard optimization problems with finite decision variables and constraints, function approximation for value estimations as well as constraint sampling are leveraged. Data-driven implementations of these algorithms are provided based on $Q$-function which enables learning the optimal value without any knowledge of the system dynamics. The performance of the approximated solutions is also verified through a weighted sup-norm bound and the Lyapunov bound. A simulation example is provided to verify the effectiveness of the presented approach.
\end{abstract}
\vspace{-0.1cm}


\IEEEpeerreviewmaketitle
\vspace{-0.25cm}
\section{Introduction}

\IEEEPARstart{M}{any} next-generation autonomous control systems such as self-driving cars will be performance-critical systems for which an acceptable performance must be guaranteed throughout their course of operation. Guaranteeing performance in the presence of uncertainties, however, is challenging. There are generally two types of uncertainties that autonomous systems encounter: 1) epistemic uncertainties which occur due to the lack of data and can be reduced by collecting data (e.g., uncertain system dynamics), and 2) aleatory uncertainties due to randomness which are inherent in the system and cannot be reduced by data collection. Examples of aleatory uncertainties for a self-driving car could be sensors measurement noise, errors in perception due to weather, etc.   

Classical optimal control methods achieve performance guarantee by optimizing the expected value of the objective function using a high-fidelity model of the system dynamics \cite {Lewis2012}-\cite{Lewis2009}. In the presence of epistemic uncertainty, a variety of adaptive dynamic programming (ADP) or reinforcement learning (RL) algorithms have been developed to learn the optimal expected value function and its corresponding control policy \cite {Lewis2009}-\cite {Wang2012}. However, basing the design on optimizing expected value of outcomes and ignoring its variance due to aleatory uncertainties can result in high-variance controllers with performance fluctuation when deployed in real systems, which is not acceptable for performance-critical systems. Instead, to take into account aleatory uncertainties, the risk-averse optimal control (RAOC) problem has been widely considered to minimize not only the mean but also the variance of the performance or cost function. The term risk averse refers to the preference on choosing a more predictable outcome (i.e., less variance)  with possibly lower expected value \cite {Ruszczy2010}-\cite {Medina2012}. Therefore, to achieve long-term optimality as well as to avoid performance fluctuation, the RAOC problem has attracted a surge of interest \cite {Gaucherand1997}-\cite {Zhu2009}. Most existing risk-averse methods ignore the epistemic uncertainty and use a high-fidelity model of the system to design a risk-averse controller for systems under aleatory uncertainty. Complex systems such as self-driving cars are under both epistemic and aleatory certainties and it is of vital importance to learn online a solution to RAOC problem through interaction with the environment and without assuming that a high-fidelity model of the system exists.

RAOC problem has been presented in different settings. Specifically, value-at-risk criteria \cite{Chapman2019}-\cite{Chapman2021} has been widely used to formulate the RAOC problem as optimizing the expectation of the cost function while assuring that its variance is below an acceptable threshold. On the other hand, entropic risk measures leverage exponential cost functions to simultaneously optimize the average cost and its variance \cite{Saldi2020}-\cite{Dvijotham2014}. To account for epistemic uncertainties, policy gradient (PG) method \cite{Sutton2018}, as an RL algorithm, has been leveraged to learn the solution to value-at-risk setting \cite{pan2019}-\cite{Mazumdar2017} and exponential utility setting \cite{Dvijotham2014}. PG algorithms explicitly parameterize the policy and update the control parameters in the direction of the gradient of the performance.  However, generally, PG methods result in nonconvex optimization problems and thus converge to a local optimum rather than a global optimum. Moreover, PG method is sensitive to fixed gradient step size, an inappropriate choice of which can easily make it divergent.  The authors in \cite{jing2020} formulated the RAOC problem as minimizing a cost-variance performance function and developed a data-driven RL algorithm for nonlinear stochastic systems. This method, however, requires solving coupled algebra Riccati equations, which are computationally expensive. Moreover, existing risk-averse learning approaches using RL require online interaction to collect data from the current policy, i.e, the samples are generated during the execution. However, in some applications, it is highly desired to turn a large dataset of samples that are collected in advance, and without further exploration, into powerful decision-making engines that minimize both the expected value of outcome and its variance.

In this paper, the following variants of RL algorithms are presented for solving the RAOC problem 1) a one-shot static convex program based RL algorithm is first developed that leverages past collected samples from the system under a variety of control policies and turns these samples into a decision-making machinery to make risk-averse decisions without requiring any online exploration. 
This is beneficial to many settings for which a large dataset of samples is available a priori and online interaction might be impractical, either because data collection is expensive or dangerous; 2) an iterative value iteration (VI) based RL algorithm is presented that solves a linear programming (LP) optimization at each iteration. Similar to the one-shot approach, it can leverage past collected data, but can also incorporate data collected from online interaction when a priori available data is not significant. A significant advantage of the VI algorithm is that it amounts to solving {an} LP optimization instead of a general convex optimization problem; and 3) an iterative policy iteration (PI) algorithm that solves a convex optimization problem at each iteration. The PI algorithm incorporates samples collected from online interaction and has the advantage of guaranteeing stochastic stability of consecutive control policies, when it starts from a stable control policy. The exact optimization problem is infinite dimensional in all three cases and their convergence to the optimal risk-averse value function is shown. To turn these optimization problems into standard optimization problems with finite decision variables and constraints, function approximation for value estimations as well as constraint sampling are leveraged, and data-driven implementations of these algorithms are provided based on $Q$-function. The performance of the approximated solutions is also verified through a weighted sup-norm bound and the Lyapunov bound. A simulation example is provided to verify the effectiveness of the presented approach.


The rest of this paper is organized as follows. In Section II, the problem formulation is given. A one-shot convex program for solving the RAOC problem is developed in Section III, and interactive LEE Bellman inequality-based VI and PI frameworks for the RAOC problem are presented in Section IV and V, respectively. Next, a simulation example is performed to verify the effectiveness of the proposed approach in Section VI. Finally, Section VII provides the conclusions.

  
\textit{Notation:} { Throughout this paper, we use $\mathbb{R}$, $\mathbb{R}_{\ge 0}$, and $\mathbb{N}$ to denote the sets of real numbers, non-negativ real numbers, and non-negative integers, respectively.} ${\mathbb{R} ^n}$ and ${\mathbb{R}^{n \times m}}$  denote the $n$ dimensional real vector space, and the $n \times m$  real matrix space, respectively. For given variables $x$ and $y$, $\big<x,y\big>$ denotes the inner product of $x$ and $y$. For a random variable $X$, $\mathbb{E}(X)$ denotes the expectation and $\mathbb{V}ar(X)=\mathbb{E}(X^2)-\mathbb{E}(X)^2$ refers the variance of $ X$, respectively. $\exp(.)$ is an exponential operator such that $\exp(x)=e^{x}$. For convenience, $\mathbb{E}_{\xi}$ refers the expectation over the random variable $\xi$. {For a function $g: X\to \mathbb{R}$, the infinity norm, the weighted 1-norm, and the weighted infinity norm are defined as $||g||_{\infty} = \sup_{x\in X}|g(x)|$,  $||g||_{1,c}=\int_{X}{|g(x)|c(x)} dx$, and $||g||_{\infty, c}={\mathop{\sup }}_{x\in X}\,c(x)|g(x)|$, respectively.} \\
\noindent \textbf{Definition 1.} The $L_{\infty}$ projection of set $V_1$ onto $V_2$ denoted as $\mathbb{P}\text{roj}_{V_2}(V_1)$ is defined as 
\begin{align}
\mathbb{P}\text{roj}_{V_2}(V_1) := \arg \min_{ V^P\in V_2}||V_1-V^P||_{\infty}. \label{eqdef1}
\end{align}
\noindent \textbf{Definition 2.} A metric space $\mathcal{S}(X)$ is called complete if every Cauchy sequence of points in $\mathcal{S}(X)$ has a limit that is also in  $\mathcal{S}(X)$.

\section{problem formulation}
Consider the system described by the following stochastic nonlinear difference equation
\begin{align}
{{x}_{t+1}}=f(x_t,u_t,\varepsilon_t)
\label{eq1}
\end{align}
 where $x_t\in {\cal X}$ and $u_t \in {\cal U}$ are the system's states and inputs, respectively, and $f: {\cal X} \times {\cal U} \times {\cal E} \to {\cal X}$ is the dynamics function. {The sets $\mathcal{X}$ and $\mathcal{U}$ are the admissible continuous state and action spaces, respectively}.  Moreover, $\left\{\varepsilon_t \in {\cal E}: t \in \mathbb{N}\right\}$ is a  stochastic process defined on a complete probability space $(\Omega, \mathscr{F}, \operatorname{Pr})$. 
 

 A nonnegative  cost $\cal{L}^{\pi}$ for a fixed control policy $\pi: {\cal X} \to  {\cal U}$ is
given by 
\begin{align}
 {\cal{L}^{\pi}}\text{=}\sum\limits_{t=0}^{\infty }\gamma^{\,t} \Big({l({{x}_{t}})+{\pi(x_t)^{T}R \pi(x_t) }} \Big),
\label{eq2}
\end{align}
where $l(.)$ is a positive definite function and $R$ is a positive definite symmetric matrix and $\gamma  \in (0,1]$ is a discounted factor. Hereafter, the subscript $t$ may be dropped for notational simplicity. The infinite-horizon risk-averse cost function $J$ can be formulated as 
\begin{align}
J(x_0,\pi)=\frac{1}{\alpha }\ln ({\mathbb{E} }[\exp (\alpha \cal{L}^{\pi})]) \label{eq3}
\end{align}
where $\alpha \in \mathbb{R}_{> 0}$ is the risk-averse factor. 

The following result has been widely used in the literature which shows that the exponential cost function \eqref{eq2} takes into account not only the expected value but also the variance of the cost \eqref{eq2}. The proof is provided in the appendix for the sake of completeness.\vspace{6pt}


\begin{lem}\label{Lemma:1}
Consider the infinite-horizon risk-averse cost function $J$ given in (\ref{eq3}). Then,
\begin{align}
J\left( {{x}_{0}},{{\pi }} \right) \approx  {\mathbb{E}}\left( \mathcal{L}_{{}}^{\pi }({{x}_{0}},{{\pi }}) \right)+\frac{\alpha }{2}\mathbb{V}ar\left( {{\mathcal{L}}^{\pi }}({{x}_{0}},{{\pi }}) \right).
\label{eq4}
\end{align}
\end{lem}
\textit{Proof.} See Appendix A. $\hfill$ $\square$  \vspace{6pt}


\begin{lem}\label{Lemma:2}
The risk-averse cost function \eqref{eq4} is continuous, non-decreasing, and convex.
\end{lem}
\textit{proof.} See \cite{Nicole18}.



\noindent \textbf{Remark 1.} Note that compared with the traditional expected cost criteria, the risk-averse objective takes the variance into consideration and the trade-off between the expectation and the variance is achieved by an appropriate selection of the risk factor $\alpha$.

We now formally state the risk-averse optimal control (RAOC) problem as follows. \vspace{6pt}\\
{
\noindent \textbf{Problem 1.}} Design an optimal control policy $\pi^{*}: {\cal X} \to {\cal U}$ that minimizes the cost {function} (\ref{eq3}). That is, find $\pi^{*}$ that solves \begin{align}
J({{x}_{0,}}\,{{\pi}^{*}})=\underset{\pi \in \Pi}{\mathop{\inf }}\,J({{x}_{0}},\pi)={{J}^{*}}({{x}_{0,}}\,\pi).\label{eq5}
\end{align} 
where $\Pi$ is the set of admissible policies.

For a given policy $\pi$, defining the value function as
\begin{align}
V^{\pi}(x) :=J({{x}_{}}\,{{,\pi}^{}}), \label{newdefvalue}
\end{align} 
the following log-expected-exponential (LEE) Bellman equation is obtained \cite{Fei2020}
\begin{align}
V^{\pi}(x)=l(x)+{{\pi(x)^{T}}R \,\, \pi(x)}+\frac{1}{\alpha }\ln ({\mathbb{E}_{x_{\pi(x)}'}}[\exp (\alpha \gamma V^{\pi}(x_{\pi(x)}'))])   \label{eq6}
\end{align}
where, in general, $x_{u}'$ is the realization of the next state starting from the state $x$ and applying the action $u$. The method that is classically implemented to solve the RAOC problem is dynamic programming (DP). DP optimality condition for the optimal value function gives the following optimal LEE  Bellman equation
\begin{align}
{{V}^{*}}({{x}}):=&\underset{\pi \in \Pi}{\mathop{\inf }} \,\, V^{\pi}(x) \nonumber \\
=&\underset{u\in {\cal U}}{\mathop{\inf }}\,\{l(x)+{{{u}^{T}}Ru}+\frac{1}{\alpha }\ln ({\mathbb{E}_{x_u'}}[\exp (\alpha \gamma {{V}^{*}}(x_u'))])\} \label{eq7}
\end{align}
After solving (\ref{eq7}) for $V^*$, the optimal controller becomes
\begin{align}
{{\pi}^{*}(x)} = \underset{{{u\in {\cal U}}_{}}}{\mathop{\arg \,\min }}\{l(x)+{{{u}^{T}}Ru}+\frac{1}{\alpha }\ln ({\mathbb{E}_{x_u'}}[\exp (\alpha \gamma {{V}^{*}}(x_u'))])\} \label{eqnew}
\end{align}
For the convenience of analysis, (\ref{eq7}) can be expressed in the form of an operator $\mathcal{T}$ as 
\begin{align}
   {{V}^{*}}(x)={\cal T}{{V}^{*}}(x), \forall x \in \cal X, \label{eq142}
\end{align}
with 
\begin{align}
   {\cal T}{{V}^{*}}(x) := {\underset{u\in \cal U}{\mathop{\inf }}}\{l(x)+{{{u}^{T}}Ru}+\frac{1}{\alpha }\ln ({\mathbb{E}_{{{x_{u}'}{{}}}}}[\exp (\alpha \gamma {{V}^{*}}(x_u'))])\}   \label{eq142}
\end{align}
 where ${\cal T}$ is the optimal LEE Bellman operator, called optimal risk Bellman operator, mapping from the space of risk averse objective functions to itself.


\subsection{ $Q$-function formulation for LEE Bellman equation }
The LEE Bellman equation (\ref{eq6}) computes the value of each state  for any admissible control policy $\pi$. Alternatively, to compute the value of each state-action pair $(x,u)$, the $Q$-function associated to control policy $\pi$ is defined as
\begin{align}
{Q^\pi }(x,u) := l(x) + {{{u^T}Ru}} + \frac{1}{\alpha }\ln ({\mathbb{E}_{{x'_{u}}}}[\exp (\alpha \gamma {V^\pi }({x'_{u}}))]). \label{eq8}
\end{align}
Comparing equations \eqref{eq6} and \eqref{eq8}, one has $V^{\pi}(x)=Q^{\pi}({x},u)$. Defining the optimal $Q$-function ${Q}^{*}({{x,u}}) := {Q^{{\pi ^*}}}(x,u)$ gives
\begin{align}
{Q^*}(x,u) =  l(x) + {{{u^T}Ru}} + \frac{1}{\alpha }\ln ({\mathbb{E}_{{x'_{u}}}}[\exp (\alpha \gamma {V^*}({x'_{u}}{\mkern 1mu} ))])   \label{Qadd}
\end{align}
with the optimal control policy given by
\begin{align}
{\pi ^*}(x) = \mathop {\arg {\mkern 1mu} \min }\limits_{u \in \cal U} {\mkern 1mu} {Q^*}({x_{}},{\mkern 1mu} {u_{}}). \label{eq10}
\end{align}
Note that 
\begin{align}
{{V}^{*}}({{x}})=\underset{u \in \cal U}{\mathop{\inf }}\,{{Q}^{*}}({{x}},u). \label{eq12}
\end{align}
{Substituting (\ref{eq12}) into (\ref{Qadd}), (\ref{Qadd}) can be expressed in the form of the optimal $Q$ risk Bellman operator $\cal F$  as }
\begin{align}
   {{Q}^{*}}(x,u)={\cal{F}} {{Q}^{*}}(x,u) 
  \label{revise1}
\end{align}
with
\begin{align}
   {\cal{F}} {{Q}^{*}}(x,u) := l(x)+{{{u}^{T}}Ru}+\underset{u' \in \cal U}{\mathop{\inf }}\,\frac{1}{\alpha }(\ln {\mathbb{E}_{x_{u}'}}[\exp (\alpha \gamma {{Q}^{*}}(x_{u}',u')))])   \label{eq9}
\end{align}
where $u'$ denotes the next action under the optimal policy $\pi^*$.

Note that the operators defined in (\ref{eq142}) and (\ref{eq9}) are nonlinear operators. In the next subsection, instead of directly solving the optimal LEE Bellman operator (\ref{eq142}) or (\ref{eq9}), easier-to-solve operators are introduced to be iterated on.

\subsection{ LEE Bellman operator and Optimal LEE Bellman operator Properties}

 
 Based on \eqref{eq6}, consider the LEE Bellman operator introduced below.
\begin{align}
   \hat{\cal T}{{V^{\pi}}}(x) := l(x)+{{{{{{u}}}}^{T}}R{{{u}}}}+\frac{1}{\alpha }\ln ({\mathbb{E}_{{{x}_{{\pi}}'}}}[\exp (\alpha \gamma {{V^{\pi}}}(x_{{\pi}}'))]),  
  \label{eq16}
\end{align}
where $u=\pi(x)$. Although the LEE Bellman operator in (\ref{eq16}) is easier to solve than (\ref{eq142}) since the infimum operation is removed, it is still a nonlinear operator, in contrast to the risk-neutral Bellman operator case, because of the nonlinearity of the exponential utility. Therefore, classical RL algorithms such as VI and PI algorithms based on temporal difference error minimization, which significantly relies on the linearity of Bellman equations, cannot be applied to solve the RAOC problem. This paper obviates this issue by formulating the RAOC problem {as convex programs in general and linear program in a special case}. Several properties of (\ref{eq16}) are investigated in this subsection, which are crucial to developing risk-averse RL algorithms for the RAOC problem in the next section. \vspace{6pt}


It is shown in the following propositions that the LEE Bellman operator $\hat{\cal T}$  and the optimal LEE Bellman operator ${\cal T}$ given in (\ref{eq16}) and (\ref{eq142}), respectively, are monotone and contractive operators.

\noindent \textbf{Definition 3.} \cite{Meyer2000,Kreyszig1978}  {Let $\mathcal{S}(X)$ be a complete space.} An operator $\mathbb{T} : \mathcal{S}(X) \to \mathcal{S}(X)$ is monotone if
\begin{align}
\big<\mathbb{T}v_1- \mathbb{T}v_2, v_1- v_2\big> \ge 0, \quad \forall v_1, v_2 \in \mathcal{S}(X). \label{eqdef2}
\end{align}

\begin{prop}\label{Proposition:1} (Monotonicity property)
The LEE Bellman operator $\hat{\cal T}$ in (\ref{eq16}) and the optimal Bellman operator {$\mathcal{T}$} in (\ref{eq142}) are monotone.
\end{prop}
\noindent \textit{Proof.} To show the monotonicity property for $\hat{\cal T}$, one needs to show that for any $V^{\pi_1}(x)$ and $V^{\pi_2}(x)$, if $V^{\pi_1}(x) \leq V^{\pi_2}(x)$, $\forall x \in {\cal X}$ then $\hat{\cal T}V^{\pi_1}(x) \leq \hat{\cal T}V^{\pi_2}(x)$, $\forall x \in {\cal X}$. Consider now two functions $V^{\pi_1}, V^{\pi_2}\in \mathcal{S}(\cal X)$. Using the fact that  $\ln(.)$ and $\exp(.)$ are injective relations (i.e., one-to-one and left-total), and $\mathbb{E}(.)$ is also injective relation over the set of strictly increasing functions, it follows from $V^{\pi_1}(x')\leq V^{\pi_2}(x')$ that
\begin{align}
   \frac{1}{\alpha }\ln ({\mathbb{E}_{x'}}[\exp (\alpha \gamma V^{\pi_1}(x'))])\le \frac{1}{\alpha }\ln ({\mathbb{E}_{x'}}[\exp (\alpha \gamma V^{\pi_2}(x'))])    \label{eqpr}
\end{align}
Adding $l(x)+{{{u}^{T}}Ru}$ to both sides of this equation yields
\begin{align}
\begin{gathered}
  l(x)+{{{u}^{T}}Ru}+\frac{1}{\alpha }\ln ({\mathbb{E}_{x'}}[\exp (\alpha \gamma V^{\pi_1}(x'))]) \le \\
  l(x)+{{{u}^{T}}Ru}+\frac{1}{\alpha }\ln ({\mathbb{E}_{x'}}[\exp (\alpha \gamma V^{\pi_2}(x'))]) \\ 
 \end{gathered}
\end{align}

\noindent
\noindent which implies that $\hat{\cal T}{V^{\pi_1}}(x) \le \hat {\cal T}{V^{\pi_2}}(x) , \,\,\, \forall x \in {\cal X}$, Therefore,
\vspace{-0.1cm}
\begin{equation}
\begin{gathered}
\big<\hat{\cal T}{V^{\pi_1}} -\hat{\cal T}{V^{\pi_2}} ,V^{\pi_1} -V^{\pi_2} \big>\ge 0 \quad \forall V^{\pi_1},V^{\pi_2}\in \mathcal{S}(\mathcal{X})
\label{eq1111}
  \end{gathered}
\end{equation}
\noindent which satisfies the condition of Definition 3. This completes the proof.  $\hfill$ $\square$
 \vspace{6pt}



Before proceeding further, the following definition and assumption are needed. \vspace{6pt}

\noindent \textbf{Definition 4.} \cite{Bertsekas2018} Let $\cal X$ be a finite-dimensional vector space. Considering a weight function $w: \cal X \to {\mathbb{ R}}$ such that $w(x) > 0, \,\, \forall x \in \cal X$, the weighted sup-norm is defined as
\begin{align}
   ||{\cal G}|{{|}_{w}} = \underset{x\in \cal X}{\mathop{\sup }}\,\frac{|{\cal G}(x)|}{w(x)}  
\end{align}

{\noindent \textbf{Assumption 1.}  There exists a constant $\rho\ge 0$ and a non-decreasing function $w: \mathbb{R}\to [1,\infty)$ such that 
\begin{align}
  l(x)+{{{u}^{T}}Ru}\le \rho w(x), \quad \forall x\in \cal X
\end{align}}
where $w(x)$ satisfies
\begin{align}
  \underset{u\in \cal U }{\mathop{\sup }}\,{\mathbb{E}_{x_{u}^{'}}}w(x_{u}^{'})\le \Delta w(x), \forall x \in \cal{X} 
\end{align}
for some constant $\Delta \in (0,\frac{1}{\gamma})$.

\begin{thm}\label{Proposition:2} (Contraction property)
Let Assumption 1 hold. The Bellman operator in (\ref{eq16}) and the optimal Bellman operator in (\ref{eq142}) are contraction mappings with respect to the weighted norm $w$. That is,
\begin{align}  
 &  ||{\cal T}{V^{\pi_2}(x)}-{\cal T}{V^{\pi_1}(x)}||_{w} \le \Delta \gamma ||{V^{\pi_1}}(x)-{V^{\pi_2}}(x)|{{|}_{w}}\,
 \label{cont1}
\end{align}

and
\begin{align}
 &  ||{\hat {\cal T}}{V^{\pi_2}(x)}-{\hat {\cal T}}{V^{\pi_1}(x)}||_{w} \le \Delta \gamma ||{V^{\pi_1}}(x)-{V^{\pi_2}}(x)|{{|}_{w}}. 
\end{align}
\end{thm}
where $\Delta \gamma \le 1$.

\textit{Proof.} See Appendix B.  $\hfill$ $\square$  \vspace{6pt}

\noindent \textbf{Remark 2.} Note that if the stage cost or reward function $l(x)+{{u}^{T}}Ru$ is bounded, then the weighted sup-norm becomes the {$\infty$} -norm and the contraction mapping is satisfied for {$\infty$}-norm. The weighted sup-norm contraction, however, captures a more general case for which the reward is only needed to be bounded by a function of the state. If the stage cost is not bounded, the {$\infty$}-norm might be unbounded. 

\begin{prop}\label{Proposition:3}
Let $V_0$ be an arbitrary value function. Then, the following properties hold.
\begin{enumerate}
\item[{1)}] 
\begin{align}
{V^\pi } = \mathop {\lim }\limits_{n \to \infty } {\mkern 1mu} {\hat{\cal T} ^n} \, {V_n}
\label{eq24}
\end{align}
\noindent for the {LEE} Bellman operator (\ref{eq16}) when the actions are generated by a fixed policy $\pi$, i.e., $u=\pi(x)$.  

\item[{2)}] 
\begin{align}
{V^* } = \mathop {\lim }\limits_{n \to \infty } {\mkern 1mu} {{\cal T} ^n} \, {V_n}
\label{eq24t}
\end{align}

\noindent for the {LEE} optimal Bellman operator \eqref{eq142}.
\end{enumerate}
\end{prop}
\noindent \textit{Proof.} Based on {Proposition \ref{Proposition:1} and Theorem \ref{Proposition:2},} the LEE Bellman operator and the optimal LEE Bellman operator are shown to have properties of monotonicity and contraction. Observing these facts, the result follows by invoking Banach's theorem \cite{Conway} which implies that {there are unique fixed-point solutions $V^{\pi}$ and $V^{*}$ of the operators.} This completes the proof. \\
\null $\hfill$ $\square$

\begin{prop}\label{Proposition:4}
A value function $V$ that satisfies the Bellman inequality ${V}\leq \hat{\cal T} V$ provides a lower bound to $V^{\pi}$. Moreover, a value function $V$ that satisfies the optimal Bellman inequality ${V}\leq {\cal T} V$ provides a lower bound to $V^{*}$.
\end{prop}
\noindent \textit{Proof.} Considering ${V}\leq \hat{\cal T} V$ and applying the operator $\hat{\cal T}$ on its both sides repeatedly, and using the results of Proposition \ref{Proposition:3} yields
\begin{align}
 \quad \quad V \leq \hat{\cal T} V \le \hat{\cal T}^2V ....\le \hat{\cal T}^{\infty}V=\underset{n\to \infty }{\mathop{\lim }} \hat{\cal T}^{n} V={{V}^{\pi}}
\label{eq25}
\end{align}
The same argument can be used also to show that ${V}\leq {\cal T} V$ provides a lower bound to $V^{*}$. This completes the proof. $\hfill$ $\square$



The main goal of the RAOC problem is to find the optimal control policy $\pi^*$ that optimizes the risk-averse cost function (\ref{eq3}), which is generally computationally intractable. To sidestep the computational complexity, three RL algorithms are developed in the next section based on different learning processes 1) one-shot static convex optimization, 2) iterative value iteration (VI) algorithm that solves an LP optimization at each iteration, and 3) iterative policy iteration algorithm (PI) that solves a convex optimization at each iteration. The convergence of the solution of all three algorithms to the solution of (\ref{eq142}) or equivalently (\ref{eq9}) is shown. The performance of the proposed algorithms is also verified through a weighted sup-norm bound and the Lyapunov bound. Based on the analysis of the properties, data-driven implementations of these algorithms are provided based on $Q$-function which enables learning the optimal value without any knowledge of the system dynamics.




%



\section{One-shot convex program for solving the RAOC problem}

\subsection{One-shot infinite-dimensional convex program}

The following one-shot optimization algorithm is presented for solving the RAOC problem and its convergence to the fixed point of $\cal T$ operator \eqref{eq142} is also shown.  \vspace{6pt}

 \noindent {\textbf{Problem 2. (One-shot infinite-dimensional convex program)}
\begin{align}
&   \underset{V\in \mathcal{S}(\cal X)}{\mathop{\max }}\,\int_{\cal X}{V(x)c(x)} dx \nonumber \\ 
& s.t. \quad V(x)\le \hat{\cal T} V(x), \quad \forall (x,u) \in \cal X \times \cal U 
 \label{value}
\end{align} 
\noindent where the state-importance weighting $c(.)$ is a finite measure on $\cal{X}$ that assigns positive weights to all open subsets of $\cal{X}$.

Note that while the operator $\hat{\cal T}$ is used in this optimization problem, since the constraints must be satisfied for all actions $u \in \cal{U}$, they are also satisfied for the infimum of $u$ as well, which means that it also satisfies the operator $\cal T$. Since the maximization finds the best lower bound for all value functions that satisfy the inequality, the optimal value function is the unique solution of this optimization.  The following corollary provides a formal proof for this claim. \vspace{6pt}
 
\begin{cor} \label{Corollary:1}
The solution to the optimization problem \eqref{value} coincides with the optimal value function $V^*$.
\end{cor}
\noindent\textit{proof.} Recall that
\begin{align}
    {{V}^{*}}(x) = & \underset{u\in \cal U}{\mathop{\inf }}\,\{l(x)+{u{{}^{T}}Ru}+\frac{1}{\alpha }\ln ({\mathbb{E}_{x_{u}'}}[\exp (\alpha \gamma {{V}^{*}}(x_{u}'))])\} \nonumber  \\ 
  & \le l(x)+{{{u}^{T}}Ru}+\frac{1}{\alpha }\ln ({\mathbb{E}_{x_{u}'}}[\exp (\alpha \gamma {{V}^{*}}(x_{u}'))])  
   \label{valuep1}
\end{align}
  $\forall (x,u)\in \cal X \times \cal U$. This implies that $V^*$ is a feasible solution to (\ref{value}). Also, for any $V\in \mathcal{S}(\cal X)$ in a feasible region of (\ref{value}), the inequality
  \begin{align}
 V(x)\le l(x)+u^{T}Ru+\frac{1}{\alpha} \ln ({\mathbb{E}_{x_u'}}[\exp (\alpha \gamma V(x_u'))])
    \label{eqproof3}
\end{align}
holds for all $(x,u) \in \cal X \times \cal U$, which implies that {
it} also holds for specific $u\in \cal U$ that minimizes $V(x_u')$. That is, it also satisfies $\cal{T}$ operator, i.e, 
$V(x)\le{{\cal T}} V(x)$. Based on Proposition \ref{Proposition:4}, this indicates that $V$ is a lower bound to $V^*$. Since the objective function in (\ref{value}) is maximized over the space of value functions, $V^*$ as the best lower bound, is the solution of the optimization. $\hfill$ $\square$

Based on Lemma \ref{Lemma:2}, the constraint set is convex and since the objective function is linear in the decision variables, Problem 2 is a convex optimization problem. This optimization problem, however, is an infinite-dimensional convex optimization problem since there are infinite-decision variables and infinite number of samples. The following subsections resolve this issue by first restricting the value function in the space of a finite family of basis functions and then sampling the constraints.


\subsection{One-shot semi-infinite-dimensional convex program}

In order to overcome the computational complexity of the infinite-dimensional state space in the optimization problem (\ref{value}), the value function is restricted in the span of a finite family of basis functions as 
\begin{align}
\hat{\mathcal{S}}({\cal X})=\{\sum\limits_{i=1}^{N}{{{h }_{i}}{{{v}}_{i}(x)}}| h \in {{\mathbb{R}}^{N}}\}
 \label{eq29}
\end{align}
where $v_i: \cal X \to \mathbb{R}$ is the finite family of basis functions of the value function and $\{h_1,...,h_N\}$ are the basis functions weights. The optimal risk-averse value function is assumed to be located in the span of the approximate space, and there is a trade-off between the computational complexity and quality of the approximation.


\noindent \textbf{Problem 3. (One-shot semi-infinite-dimensional convex program) } 
\begin{align}
&\underset{\hat{V}\in \hat{\mathcal{S}}({\cal X})}{\mathop{\max }}\,\int_{\cal X}{\hat{V}(x)c(x)} dx \nonumber \\ 
& s.t. \quad \hat{V}(x)\le \hat{\cal T}\hat{V}(x),   \forall (x,u) \in \cal X \times \cal U   
 \label{eq30n}
\end{align}

Note that the number of decision variables is now finite (i.e, the size of the weight vector $h=[h_1,...,h_N]$), but the number of constraints is still infinite, which makes the optimization problem a semi-infinite convex program. 

The following lemma shows that Problem 3 results in the minimization of a certain weighted norm, which suggests that the weights $c(x)$ impose a trade-off in the quality of the value function approximation across states. \vspace{6pt}

\begin{lem}\label{Lemma:3}
Let $V^{*}$ be the solution to the exact convex program \eqref{value}. Then, $\hat{V}$ solved in (\ref{eq30n}) approximate $V^{*}$ if and only if
\begin{align}
&   \underset{\hat{V}\in \hat{\mathcal{S}}({\cal X}) }{\mathop{\min }}\,||{{V}^{*}}-\hat{V}||_{1,c(x)} \nonumber \\ 
&   s.t.\quad \hat{V}(x)\le \hat{\cal T}\hat{V}(x),  \quad \forall (x,u) \in \cal X \times \cal U   
 \label{eq31t}
\end{align}
\end{lem}
\textit{Proof.} See Appendix C.  $\hfill$ $\square$ \vspace{6pt}


To investigate the performance bound of the proposed scheme for the RAOC problem, inspired by \cite{Farias2003, Farias2004, Beu}, which performed the analysis on LP-based risk-neutral approximate dynamic programming,  the error bounds due to approximating the risk-averse value function (i.e., the approximation error) are now investigated. In contrast to \cite{Farias2003, Farias2004, Warrington2019}, however, weighted sup-norm is used in our analysis, because of the weighted sup-norm contraction results for risk-averse case in Theorem 1. The optimal value function found by the exact optimization  (\ref{value}) (called $V^{*}$ here) is now compared with the value function solution to the optimization Problem 3. Let $V^P$ be the projection of ${V}^{*}$ on $\hat{\mathcal{S}}({\cal X})$. Even if the basis functions are chosen such that their span is relatively close to $V^{*}$ (i.e., $||{{V}^{*}}-V^P\|{{|}_{w }}$ is small), the approximate value function found by solving Problem 3 (called ${\hat{V}}^{*}$ here) might not be close to $V^{*}$ in general. 


\begin{thm}\label{Theorem:1} (weighted sup-norm bound)
Let ${{\hat{V}}}^{*}$ be the approximate value function solution to  (\ref{eq30n}), and  $V^*$ be the solution to (\ref{eq7}). Then, 
\begin{align}
    &  \int\limits_{\cal X}\frac{{|{{V}^{*}}-}{{{\hat{V}}\,}^{*}}|}{w(x)}c(x)dx \le \frac{2}{1-{{{\Delta \gamma } }^{}}}||{{V}^{*}}-V^P{{||}_{w }}.
\label{eq46}
\end{align}
\end{thm}
\noindent \textit{Proof}. See Appendix D.  $\hfill$ $\square$ 



Before proceeding to the Lyapunov bound analysis, the following lemmas and definition are required.

\noindent \textbf{Definition 4.} $\mathbb{M}$ operator is defined as follows
\begin{align}
  \mathbb{M}(V^{\pi}(x)) :=\underset{u\in {\cal U}}{\mathop{\max }}\,\mathbb{E}[V^{\pi}(f(x,u,\varepsilon))]
\label{eq47}
\end{align}
which implies the worst case of the future risk-averse value function. Moreover, based on (\ref{eq47}), the Lyapunov function $V$ is defined as the one whose Lyapunov factor $\Theta_V$ satisfies the following condition
\begin{align}
  \Theta_V =\underset{x\in {\cal X}}{\mathop{\max }}\, \frac{\gamma \mathbb{M}(V^{\pi}(x))}{V^{\pi}(x)} < 1  
\label{eq48}
\end{align} 

\begin{lem}\label{Lemma:4}
For any $g_1 (x)$, $g_2(x)$ $: {\cal X} \to \mathbb{R}$, the following holds
\begin{align}
|{\cal T}g_1(x)-{\cal T}g_2(x)|\le \gamma \mathbb{M}(|g_1(x)-g_2(x)|) 
\end{align}
\end{lem}
\noindent \textit{Proof.} See Appendix E. $\hfill$ $\square$  \vspace{6pt}

\begin{lem}\label{Lemma:5}
For any positive function $J(x)$ and any function $V(x)$, let $\delta :=||V^*-V||_{\infty, \frac{1}{J}}$, then following holds
\begin{align}
{\cal T} V(x)\ge V(x)-\delta J(x)-\gamma \delta \mathbb{M}(J(x)) 
\end{align}
\end{lem}
\noindent \textit{Proof}: See Appendix F.  $\hfill$ $\square$  \vspace{6pt}

\begin{lem}\label{Lemma:6}
Let $V_L$ be the Lyapunov function defined in (\ref{eq48}), $V(x)$ be any value function, and
\begin{align}
 \overline{V}=V-||{{V}^{*}}-V|{{|}_{\infty ,\frac{1}{V_L}}}(\frac{2}{1-{{\Theta }_{V_L}}}-1)V_L 
 \label{eq:44lema}
\end{align}
 Then, $\overline {V}$ is a solution to (\ref{eq30n}), i.e., $\overline{V} \le \hat{\cal T} \overline{V}$.
\end{lem}
\noindent \textit{Proof}: See Appendix G. \vspace{6pt}

\begin{thm}\label{Theorem:2} (Lyapunov bound)
Let $V^*$ be the solution to (\ref{eq7}), and $\hat{V}^*$ be the solution to (\ref{eq30n}). If $\hat{V}^{+}>0$ satisfies $\gamma \mathbb{M} \hat{V}^{+} < \hat{V}^{+}$, then 
\begin{align}
||V^*-{{\hat{V}}\,}^{*}||_{1,c(x)}\le \frac{2||{{\hat{V}}^{+}}|{{|}_{1,c(x)}}}{1-\Theta _{{{\hat{V}}^{+}}}^{}}\underset{\hat{V}\in \hat{\mathcal{S}}({\cal X})}{\mathop{\min }}\,||{{V}^{\text{*}}}\text{-}\hat{V}|{{|}_{\infty ,\frac{1}{{{\hat{V}}^{+}}}}}.   
\label{eq49}
\end{align}
\end{thm}
\noindent \textit{Proof.} See Appendix H.  $\hfill$ $\square$ 

An advantage of this Lyapunov bound over the weighted sup-norm bound is that  a mechanism for selection of the relevance weighting, which is showed up in the right-hand side of the Lyapunov bound, can be devised.

\subsection{Model-free one-shot standard convex program}
To provide tractable standard convex programs for solving the one-shot optimization problem (with a finite number of decision variables and samples), a sampling-based one-shot algorithm is now presented. While a sample-based standard convex optimization algorithm can be developed by sampling constraints from the optimization Problem 3, it would result in a model-based greedy policy learning. That is, after finding the approximate optimal value function $\hat V^*$ from solving sampled-based version of Problem 3, finding the greedy approximate optimal control policy by solving
\begin{align}
{\hat{\pi}}({{x}})=\underset{{{u}_{}\in {\cal U}}}{\mathop{\arg \,\min }}\{l(x)+{{{u}^{T}}Ru}+\frac{1}{\alpha }\ln ({\mathbb{E}_{x_{u}'}}[\exp (\alpha \gamma {{{\hat V}^{*}}}(x_u'))])\}
\label{eq32}
\end{align}
requires the complete knowledge of the system dynamics. However, since the convergence properties of Problem 1 and performance bounds for Problem 3 also hold if $\hat {\cal T}$ operator is replaced with $\hat {\cal F}$ operator, one can replace value-based equation (\ref{eq30n}) with the following optimization based on $Q$ function
\begin{align}
&   \underset{\hat{Q}^i\in \hat{\mathcal{S}}({\cal X},{\cal U})}{\mathop{\max }}\,\int_{{\cal X} \times {\cal U}}{\hat{Q}^{} (x,u)c(x,u)dxdu} \nonumber  \\ 
& \hat{Q}^{}(x,u)\le l(x)+{{(u)^{T}}R{u^{}}}+\frac{1}{\alpha }\ln ({\mathbb{E}_{x_{u}}^{'}}[\exp (\alpha \gamma \hat{Q}^{}(x_{{{u}}}^{'},u') )]) \nonumber \\
&  \forall (x,u,u') \in {\cal X} \times {\cal U} \times {\cal U}
 \label{eqQoper}
\end{align}
Similar to the value function approximation, the space of $Q$-functions is also limited to a set of basis functions defined as
\begin{align}
\hat{\mathcal{S}}({\cal X} \times {\cal U})=\{\sum\limits_{i=1}^{N}{{{\beta }_{i}}{{{q}}_{i}(x,u)}}|\beta \in {{\mathbb{R}}^{N}}\}
 \label{Qfun}
\end{align}
where  $q_i: \cal X \times \cal U \to \mathbb{R} $ is the finite family of basis functions of the $Q$-functions, with $\beta={\{\beta_1, \beta_2, ..., \beta_N\}}$ as the weights. The optimal risk-averse $Q$-function is assumed to be located in the span of the approximate space, and there is a trade-off between the computational complexity and quality of the approximation. Note that the approximate quality relies on several factors (e.g., the number of basis functions, weighting factor $c(.)$), and the optimal solution $Q^*$ is assumed to be in the span of the approximate space  $\hat{\mathcal {S}}(\cal X, \cal U)$. The details of data collection and implementation of this data-based algorithm are postponed to Section V. A since Algorithm 1 and Algorithms 3 and 4 that will be presented later for VI and PI have the same data collection strategies. 

Algorithm 1 uses this optimization for a model-free implementation of the one-shot optimization.


\begin{algorithm}[H]
	\caption{One-shot sample-based standard convex optimization algorithm.}
	\SetAlgoLined
	\KwResult{The optimal risk-averse objective function $\hat{Q}^*$.}
	\textbf{Input:} {The basis functions; the noise $\kappa$ in control input $u$ and the noise $\upsilon$ in state $x$; the number of sample tuples $N$, $Z$ and number of control policy $O$.} \\
	\begin{enumerate}
		\item [1:] \textbf{Data collection:} Select different control policies $\pi=\{ \pi_1, \pi_2, ..., \pi_o \}$. For each control policy $\pi_o \in \pi$, generate state-action tuples ${{\cal D}_s}_{\pi_o}^N:={\{(x^s,\pi_o(x^s),(x_{{{\pi_o^i}}}^{'})^s, \pi_o((x_{{{\pi_o^i}}}^{'})^s))\}}_{s=1}^N$ randomly with noise $\kappa$ and $\upsilon$ in the control input $\pi_o$ and state $x$ where $(x_{{{\pi_o^i}}}^{'})^s\sim f(x^s,\pi_o(x^s),\varepsilon_t)$. 
		
		\item [2:] \textbf{Function evaluation:} Solve
		\begin{align}
			&   \underset{\hat{Q}^\in \hat{\mathcal{S}}({\cal X},{\cal U})}{\mathop{\max }}\,\int_{{\cal X\times \cal U}}{\hat{Q}^{} (x,u)c(x,u)dxdu} \nonumber  \\ 
			& \hat{Q}(x^s,\pi_o(x^s))\le l(x^s)+{{(\pi_o(x^s))^{T}}R{\pi_o(x^s)}}\nonumber \\ &+\frac{1}{\alpha }\ln (\frac{1}{Z}\sum_{i=1}^Z{{}}[\exp (\alpha \gamma \hat{Q}(x_{{{{\pi_o}^i}}}^{'})^s,\pi_o((x_{{{{\pi_o}^i}}}^{'})^s)) )]) \nonumber \\
			&  \forall {\{(x^s,\pi_o(x^s), (x_{{{\pi_o^i}}}^{'})^s, \pi_o((x_{{{\pi_o^i}}}^{'})^s)  )\}} \in  {{\cal D}_s}_{\pi_o}^{N}  
			\label{eqQoper}
		\end{align}
	\end{enumerate}
\end{algorithm}

Letting ${\hat{Q}^{{}}}$ be the solution to Algorithm 1, the model-free greedy policy is found by
\begin{align}
{{\pi}^{}}=\underset{u \in {\cal U}}{\mathop{\arg \min }}\,{\hat{Q}^{{}}}(x,u)    \label{qupd}
\end{align}




\noindent \textbf{Remark 3.} Note that a significant advantage of this algorithm is that the samples can be collected in advance and no online interaction or consecutive samples are required to be collected, which provides a machinery to turn large data sets into decision-making mechanisms. This algorithm, however, requires solving a convex optimization problem. The next section shows how an iterative VI algorithm can be leveraged to solve an LP optimization at each iteration.

\section{Interactive LEE Bellman inequality based VI framework. }

Recalling $\mathcal{T}$ operator developed in (\ref{eq142}), the following VI algorithm solves an LP optimization at each iteration. \vspace{6pt}

\noindent {\textbf{Problem 4. (Interactive infinite-dimensional LP-based VI algorithm)} For each $k\in {\{0,...,N-1\}}$,  $V^k$ is the solution to the following LP optimization problem
\begin{align}
&   \underset{V^{k+1}\in \mathcal{S}({\cal X})}{\mathop{\max }}\,\int_{{\cal X}}{V^{k+1}(x)c(x)} dx \nonumber \\ 
& s.t. \quad V^{k+1}(x)\le {\cal T} V^k(x), \, \forall x \in {\cal X}
 \label{eqVI1}
\end{align}

\begin{prop}\label{Proposition:5}
The solution for (\ref{eqVI1}) is the solution to the following dynamic recursion involving value function $V^k: {\cal X} \to \mathbb{R}$ to $k=0,1,...,N-1$
\begin{align}
{{V}^{k+1}}({{x}}):=&\underset{\pi \in {\Pi}}{\mathop{\inf }} \,\, V^{k}(x) \nonumber \\
=&\underset{u \in {\cal U}}{\mathop{\inf }}\,\{l(x)+{{{u}^{T}}Ru}+\frac{1}{\alpha }\ln ({\mathbb{E}_{x_u'}}[\exp (\alpha \gamma {{V}^{k}}(x_u'))])\} \label{eqVI2}
\end{align}
\end{prop}

\noindent \textit{Proof}. Let $J^*$ be the solution to (\ref{eqVI2}) and assume that there exists a subspace $ \mathcal{B}_c \in {\cal X}$ such that $\forall x'\in \mathcal{B}_c$, $J^*(x')\ne V^{k+1}$. Since $J^*(x')$ is feasible, $J^*(x')< V^{k+1}(x')$. Then increasing $J^*(x')$ on $\mathcal{B}_c$ until $J^*(x')= V^{k+1}(x')$ increases $\int_{\cal X}{J^*(x)c(x)} dx$. This concludes that $J^*(x')$ could be optimal unless $J^*(x')= V^{k+1}(x'), \forall x\in \mathcal{B}_c$. $\hfill$ $\square$ 

Note that ${{V}^{k+1}}(x)\ge {\cal T}{{V}^{k}},\,\,\,\forall x\in {\cal X} \Leftrightarrow {{V}^{k+1}}\ge \hat{\cal T}{{V}^{k}},\,\,\forall (x,u)\in {\cal X} \times {\cal U}$. Therefore, Problem 4 is equivalent to 
\begin{align}
&   \underset{V^{k+1}\in \mathcal{S}({\cal X})}{\mathop{\max }}\,\int_{{\cal X}}{V^{k+1}(x)c(x)} dx \nonumber \\ 
& s.t. \quad V^{k+1}(x)\le \hat{\cal T} V^k(x), \, \forall (x,u)\in {\cal X} \times {\cal U}
 \label{eqVI3}
\end{align}
which is an infinite-dimensional linear optimization problem both in cost and constraints over the infinite-dimensional space of $V^{k+1}$ and with infinite number of constraints. $\hfill$ $\square$ \vspace{6pt}

\begin{prop}\label{Proposition:6}
Given an admissible initial value function $V^0$, the interactive problem (\ref{eqVI1}) converges to optimal value $V^*$.
\end{prop}

\noindent \textit{Proof.} The proof follows Proposition \ref{Proposition:5}. Since VI algorithm implement $\cal{T}$ operator in each iteration, the sequence ${\{V^0, V^1, ...V^n\}}$ converges to the fixed-point solution $V^*$. This completes the proof.  $\hfill$ $\square$ \vspace{6pt}

\noindent \textbf{Remark 4.} Note that even though the risk-averse cost function is convex, the VI algorithm uses a recursive inequality for which in its right-hand side the decision variables from the previous iteration are used (which are known), and thus inequalities are linear in decision variables (which appear in the left-hand side of inequalities). 

Based on (\ref{eqVI3}), the data-driven VI algorithm with finite-dimensional LP is now developed. Using a similar terminology used for one-shot optimization, $Q$-function approximation and $Q$-Bellman operator are used instead of value function approximation to provide a model-free data-based implementation. Algorithm 2 provides the details of sampling-based model-free $Q$-learning VI algorithm.

\begin{algorithm}[H]
\caption{Data-driven sample-based standard LP-based iterative VI algorithm.}
\SetAlgoLined
\KwResult{The optimal risk-averse objective function $\hat{Q}^*=\hat{Q}^N$.}
\textbf{Input:}{ Basis centers ${\{c_1,c_2,...c_M\}}$ and basis functions, the noise $\kappa$ in control input $u$ and the noise $\upsilon$ in state $x$; the number of sample tuples $N$, $Z$ and number of control policy $O$; the initial value function $\hat{Q}^0$} and threshold value $\xi$. \\
\textbf{Data Collection:} Select different control policies $\pi=\{ \pi_1, \pi_2, ..., \pi_o\}$. For each control policy $\pi_o \in \pi$, generate state-action tuples  ${{{\cal D}_s}_{\pi_o}}^N:={\{(x^s,\pi_o(x^s), (x_{{{{\pi_o}^i}}}^{'})^s, \pi_o((x_{{{\pi_o^i}}}^{'})^s)  )\}}_{s=1}^N$  randomly with noise $\kappa $ and $\upsilon$ in the control input $\pi_o$ and state $x$ where $(x_{{{\pi_o^i}}}^{'})^s\sim f(x^s,\pi_o(x^s),\varepsilon_t)$.\\
\While{$\null$ $\left| {{{\hat Q}^{n}} - {{\hat Q}^{n-1}}} \right| \ge \xi$ }{
    $\null$    $\quad$  \textbf{\textup{for}} all $(x^s,\pi_o(x^s), (x_{{{\pi_o^i}}}^{'})^s, \pi_o((x_{{{\pi_o^i}}}^{'})^s)  )$ $\textup{\textbf{do}}$\\
  $\null$ $\quad$ $\quad$ Evaluate $E(x^s,\pi_o(x^s))=\hat{\cal F}\hat{Q}^n(x^s,\pi_o(x^s))$\\
     $\null$ $\quad$ $\textup{\textbf{end}}$ \\
    evaluate $\hat{Q}^{n+1}$ by solving optimization problem 
\begin{align}
& \underset{\hat{Q}^{n+1}\in \hat{\mathcal{S}}({\cal X})}{\mathop{\max }}\,\int_{{\cal X\times U}}{\hat{Q}^{n+1}(x,u)c(x,u)} dxdu \nonumber \\ 
&  s.t. \quad \hat{Q}^{n+1}(x^s,\pi_o(x^s))\le E(x^s,\pi_o(x^s)), \nonumber\\  
&  \forall {\{(x^s,{\pi}_o(x^s), (x_{{{\pi_o^i}}}^{'})^s, \pi_o((x_{{{\pi_o^i}}}^{'})^s)  )\}} \in  {{\cal D}_s}_{\pi_o}^N 
 \label{eqVI4}
\end{align}
 $n=n+1$ } 
Measure the state $x^s \in {\cal X}$ \\
Sample $N_s$ points $u^s$ uniformly from ${\cal U}$\\
 \textbf{\textup{for}} all $u^s$ \textup{\textbf{do}}\\
  Evaluate $b(u^s)=\hat{\cal{F}}\hat{Q}^{N}(x^s,u^s)$\\
 \textup{\textbf{end}}\\
 Calculate $u^N= \arg \,{\underset{u^s}{\mathop{\min }}}\,b(u^s)$\\
 Apply $u^N$ to the system.
\end{algorithm}




\noindent \textbf{ Remark 5.} Notice that Algorithm 2 solves the RAOC problem in a recursive way. With the initial value function and sampling data, the improved value functions ${\{\hat{V}^1, \hat{V}^2,...\hat{V}^n\}}$ are computed successively.

\section{PI algorithm for solving RAOC problem}

In this section, a convex-program-based PI algorithm is presented for solving the RAOC problem. The risk-averse PI algorithm relies on the policy evaluation step that leverages the LEE Bellman inequality to evaluate the value function corresponding to a control policy and the policy improvement step that finds an improved policy at each iteration. The implementation of PI using a convex program is discussed in Section V.A. 

The policy evaluation step is formulated as follows. \vspace{6pt}


\noindent {\textbf{Problem 5. (Infinite-dimensional convex program for policy evaluation)} For a fixed control policy $u=\pi(x) $, solve
\begin{align}
&   \underset{V\in \mathcal{S}({\cal X})}{\mathop{\max }}\,\int_{{\cal X}}{V(x)c(x)} dx \nonumber \\ 
& s.t. \quad V(x)\le \hat{\cal T} V(x), \, \forall x \in \cal X
 \label{eq26}
\end{align}
where $\hat{\cal T}$ is defined in (\ref{eq16}).  \vspace{6pt} 

\begin{prop}\label{Proposition:7}
The solution to (\ref{eq26}) is equivalent to the solution to (\ref{eq8}), and is the {fixed-point} solution of (\ref{eq16}). 
\end{prop}

\noindent \textit{Proof.} Based on Proposition \ref{Proposition:4}, any feasible solution of (\ref{eq26}) satisfies $V\le V^{\pi}$. Therefore, maximization in (\ref{eq26}) finds the best lower bound, and thus $V^{\pi}$ is the unique solution to this optimization, which is also the fixed point of (\ref{eq16}) as shown in Proposition \ref{Proposition:3}. This completes the proof.  $\hfill$ $\square$ \vspace{6pt}

\noindent \textbf{Remark 6.} Note that the feasible region in (\ref{eq26}) can be enlarged by applying iteratively $\hat{\cal T}$ operator inequality. A value function satisfying $V\le \hat{\cal T}^{\mathnormal{M}} V$ with $M \in \mathbb{R}$ is an under-estimator of $V^{\pi}$. Notice that the constraint in (\ref{eq26}) is a special case when $M=1$ and without loss of generality, $M$ is set to $1$ in the sequel. It follows from Proposition \ref{Proposition:3} that $\hat{\cal T}^M V\xrightarrow{M\to \infty} V^{\pi}$ which also establishes the following program
\begin{align}
 &  \underset{V\in \mathcal{S}({\cal X})}{\mathop{\max }}\,\int_{{\cal X}}{V(x)c(x)} dx \nonumber \\ 
 & s.t. \quad V(x)\le \hat{\cal T}^M V(x), \, \forall x \in {\cal X}, u=\pi(x)  
 \label{eq266}
\end{align}

\noindent \textbf{Remark 7.}  The difference between the optimization (\ref{eq26}) and optimization \eqref{value} is that the control actions $u$ are relaxed in the constraints of  \eqref{value} and can be chosen arbitrary and does not need to follow any specific policy, while the control actions in (\ref{eq26}) {follow} the policy under evaluation.

Based on the finite approximation of the value function space (\ref{eq29}), an approximation version of (\ref{eq26}) is {formulated} as Problem 6. \\ 

\noindent \textbf{Problem 6. (Semi-infinite convex program based PI) } 
For a fixed control policy $\pi$, solve 
\begin{align}
&\underset{\hat{V}\in \hat{\mathcal{S}}({\cal X})}{\mathop{\max }}\,\int_{{\cal X}}{\hat{V}(x)c(x)} dx \nonumber \\ 
& s.t. \quad \hat{V}(x)\le \hat{\cal T}\hat{V}(x),   \forall x \in {\cal X}, u=\pi(x)  
 \label{eq30}
\end{align}

\begin{lem}\label{Lemma:7}
$\hat{V}$ solved in (\ref{eq30}) approximate $V^{\pi}$ if and only if
\begin{align}
&   \underset{\hat{V}\in \hat{\mathcal{S}}({\cal X}) }{\mathop{\min }}\,||{{V}^{\pi}}-\hat{V}||_{1,c(x)} \nonumber \\ 
&   s.t.\quad \hat{V}(x)\le \hat{\cal T}\hat{V}(x),  \quad \forall (x,u) \in {{\cal{X}} \times {\cal{U}}}, u=\pi(x)  
 \label{eq31}
\end{align}
\end{lem}
\noindent \textit{Proof.} The proof is similar to that of Lemma \ref{Lemma:3} and only the operator $\cal{T}$ in Lemma \ref{Lemma:3} is now replaced by the operator $\cal {\hat T}$ and thus $V^*$ is replaced by $V^{\pi}$ in the poof of Lemma \ref{Lemma:3}. $\hfill$ $\square$ \vspace{6pt}

Given the value function $V$ approximated from (\ref{eq26}) or (\ref{value}), the corresponding improved control policy or optimal control policy could be computed by 
\begin{align}
{\hat{\pi}}({{x}})=\underset{{{u}_{}\in {\cal U}}}{\mathop{\arg \,\min }}\{l(x)+{{{u}^{T}}Ru}+\frac{1}{\alpha }\ln ({\mathbb{E}_{x_{u}'}}[\exp (\alpha \gamma {{{\hat V}^{\pi}}}(x_u'))])\}
\label{eq32}
\end{align}
which requires the dynamic knowledge of the system. The following model-based algorithm is presented based on value function learning. A $Q$-function-based version of the value estimation is presented later to obviate the requirement of the system dynamics while only leveraging the samples collected from the experiments or simulations.

\begin{algorithm}[H]
\caption{Model-based PI algorithm.}
Until convergence, repeat the following steps:\\
\begin{enumerate}
\item [1:] \textbf{Policy evaluation step:} For the policy $\pi^i$, solve for $\hat{V}^{\pi^i}$ using
\begin{align}
& \underset{\hat{V}^i\in \hat{\mathcal{S}}({\cal X})}{\mathop{\max }}\,\int_{{\cal X}}{\hat{V}^i (x)c(x)dx} \nonumber \\ 
& s.t. \,\hat{V}^i(x)\le l(x)+{{{({{u}})}^{T}}R{{u}}}+\frac{1}{\alpha }\ln ({{\mathbb{E}}_{x_{u}^{'}}}[\exp (\alpha \gamma {\hat{V}^i_{{{}}}}(x_{{{u}}}^{'}))]) \nonumber \\
& \quad \forall (x,u) \in {\cal X} \times {\cal U}, u=\pi^i(x)
 \label{eq333}
\end{align}
\item [2:] \textbf{Policy improvement step:} Update the control policy $\pi^{i+1}$ as:
\begin{align}
{\pi ^{i + 1}} = \mathop {\arg {\mkern 1mu} \min }\limits_{{u_{}} \in {\cal U}} \{ l(x) + {{{u^T}Ru}} + \frac{1}{\alpha }\ln ({{\mathbb{E}}_{x_{u}^{'}}}[\exp (\alpha \gamma {\hat V^i}({x_{{u^\prime }}})])\}
\label{eq34}
\end{align}
\end{enumerate}
\end{algorithm}

\begin{lem}\label{Theorem:3} (Stochastic Asymptotic Stability)
Consider the stochastic discrete-time system (\ref{eq1}) and let $V(x): {\cal{X}} \to \mathbb{R}$ be a continuous positive definite function. Define the set $P_{\lambda}(V, \lambda) := \{x \in {\cal{X}}:
0 \le V (x) \le \lambda \}$ for some $\lambda \in \mathbb{R}_{\ge 0}$. Assume that
$ \mathbb{E} \big[V(x(t+1))|x(t)\big]- V(x(t)) \le 0, \, \forall t$. Then, for any initial condition $x_0 \in P_{\lambda}$, $x(t)$ converges to the origin {with a probability of at least} $\mathscr{P}(V(x),\lambda) := 1 - {V(x_0)}  \big/ \lambda$.
\end{lem}
\noindent \textit{Proof.} See \cite{Kushner65}.  $\hfill$ $\square$ 

\begin{thm} \label{Corollary:2}
Let the control policies and value functions obtained from the iterative PI Algorithm 3 at iteration $i$ be $\pi^i$ and ${V}^{\pi^i}$, respectively, and $\gamma = 1$. Then, if $\pi^1$ is asymptotically stable, for any initial condition $x_0 \in P_{\lambda}({V}^{\pi^i},\lambda)$, the control policy $\pi^i$ for $i = 2,3,...$ are also asymptotically stabilizable with a probability at least $\mathscr{P}(V^{{\pi ^i}},\lambda)$.
\end{thm}
\noindent \textit{Proof.} Proposition \ref{Proposition:5} shows that for a given policy the policy evaluation step (\ref{eq333}) is equivalent to the Bellman operator. To show that $\pi^{i+1}$ is stabilizing, we consider the value function ${V}^{\pi^{i} }$ as a Lyapunov candidate function. Before proceeding, note that using Jensen inequality \cite{Andrei2007} for the Bellman equation (\ref{eq6}), one has
\begin{align}
& {V}^{\pi^i}(x)=l(x)+(\pi^{i}(x))^{T}R \pi^i(x) +\frac{1}{\alpha }\ln (\mathbb {E}[\exp (\alpha  {V}^{\pi^i}({x}_{\pi^i(x)}')]) \nonumber \\
& \quad \qquad  \ge l(x)+{{\pi^i(x)^{T}}R \,\, \pi^i(x)}+ \mathbb{E} [{V}^{\pi^i}({x}_{\pi^i(x)}')]
\label{lyap1}
\end{align}
Then, 
\begin{align}
\mathbb{E} [{V}^{\pi^i}({x}_{\pi^{i+1}(x)}')]  & \leq  \mathbb{E} [{V}^{\pi^i}({x}_{\pi^{i+1}(x)}')]-  \mathbb{E} [{V}^{\pi^i}({x}_{\pi^{i}(x)}')]+ {V}^{\pi^i }(x) \nonumber \\
&\quad  -l(x)-{{(\pi^i(x))^{T}}R \,\, \pi^i(x)} \nonumber \\
&\leq {V}^{\pi^i }(x)
\label{lyap2}
\end{align}
where the first inequality is obtained from (\ref{lyap1}) and the second inequality is obtained based on the facts that $l(x)+{{\pi^i(x)^{T}}R \,\, \pi^i(x)} \ge 0$ and  the value function improves at every iteration (i.e., $\mathbb{E} ({V}^{\pi^i}({x}_{\pi^{i+1}(x)}')-\mathbb{E} ({V}^{\pi^i}({x}_{\pi^{i}(x)}') \le 0$).
Therefore, based on Lemma \ref{Theorem:3} one can conclude that for any initial condition $x_0 \in P_{\lambda}({V}^{\pi^i},\lambda)$, the control policy $\pi^i$ for $i = 2,3,...$ are also asymptotically stabilizable with probability at least $\mathscr{P}(V^{{\pi ^i}},\lambda)$. $\hfill$ $\square$



The following Proposition shows convergence of infinite-dimensional PI algorithm to the optimal policy when no function approximation is used in Algorithm 3. For the semi-infinite PI with function approximation, Theorems 5 and 6 provide the performance bounds for the case where function approximation is used. 
\begin{prop}\label{Proposition:8}
Algorithm 3 converges to the unique optimal solution, which is the fixed point of ${\cal T}$ in (\ref{eq142}) for the infinite-dimensional PI.
\end{prop}
\noindent \textit{Proof.}  We first show that $V^{\pi^{i+1}} \le V^{\pi^{i}}$. Based on the policy improvement step, one has
\begin{align}
{{V}^{\pi^i }}(x)\,\,\ge {{Q}^{\pi^i }}(x,\pi^{i+1}(x)), \forall x\in \cal X 
\end{align}
which yields
\begin{align}
 &  {{V}^{\pi^i }}(x)\,\,\ge {{Q}^{\pi^i }}(x,\pi^{i+1}(x))  =l(x)+{\pi^{i+1}(x) {^{T}} R \,\, \pi^{i+1}(x)} +\frac{1}{\alpha } \times \nonumber \\
 & \ln ({\mathbb{E}_{x_{\pi^{i+1}(x)}'}}[\exp (\alpha \gamma {{V}^{\pi^i }}(x_{\pi^{i+1}(x)}'))]) \ge l(x)+{\pi^{i+1}(x) {^{T}} R  \pi^{i+1}(x)} \nonumber  \\
& +\frac{1}{\alpha }\ln ({\mathbb{E}_{x_{\pi^{i+1}(x)}'}}[\exp (\alpha \gamma {{Q}^{\pi^i }}(x_{\pi^{i+1}(x)}',\pi^{i+1}(x_{\pi^{i+1}(x)}')))]) \nonumber \\ 
&  =\frac{1}{\alpha }\ln ({\mathbb{E}_{x_{\pi^{i+1}(x)}'}}[\exp (\alpha \gamma [{\pi^{i+1}(x_{\pi^{i+1}(x)}') {^{T}} R \,\, \pi^{i+1}(x_{\pi^{i+1}(x)}')} \nonumber \\ 
& + l(x_{\pi^{i+1}(x)}')+\frac{1}{\alpha }\ln ({\mathbb{E}_{x'}}[\exp (\alpha \gamma {{V}^{\pi^i }}((x_{\pi^{i+1}(x)}')'_{ \pi^{i+1}(x_{\pi^{i+1}(x)}')}))])])]) \nonumber \\
& +l(x)+{\pi^{i+1}(x) {^{T}} R \,\, \pi^{i+1}(x)} \nonumber  \\
& \qquad  \vdots  \nonumber \\ 
&  ={{V}^{^{\pi ^{i+1}}}}(x) 
\end{align}
Therefore, one has $V^{\pi^{1}}\ge V^{\pi^{2}} \ge ...\ge V^{\pi^{n}}$, i.e., the sequence is decreasing, and since the value functions are also bounded, it converges. On convergence, $V^{\pi^{n}}=V^{\pi^{n+1}}=V^{\pi^{*}}=V^*$. Since $\pi^{n+1}$ is the greedy policy solved from $V^{\pi}$ using (\ref{eq34}), the following holds
\begin{align}
   {V^{{\pi ^{n + 1}}}}(x) & = \mathop {\min }\limits_{u \in {\cal U}} \{ l(x) + {{{u^T}Ru}} + \frac{1}{\alpha }\ln ({\mathbb{E}_{{x_u}}}[\exp (\alpha \gamma {V^{{\pi ^n}}}({x_{u'}}))])\}    \nonumber \\
  & =\underset{u\in {\cal U}}{\mathop{\min }}\{l(x)+{u{{}^{T}}Ru}+\frac{1}{\alpha }\ln ({\mathbb{E}_{x_u}}[\exp (\alpha \gamma {{V}^{\pi^{n+1}}}(x_{u}'))])\}    \label{PICON}
\end{align}
which is optimal LEE Bellman equation (\ref{eq142}). Since $\pi^{n+1}$ satisfies LEE Bellman optimal equation (\ref{eq142}), $\pi^{n+1}=\pi^*$ holds, meaning that both $\pi^{n}$ and $\pi^{n+1}$ are optimal policies. In other words, policy improvement yields a strictly better policy unless the original policy is already optimal. This completes the proof.  $\hfill$ $\square$

\begin{thm} \label{Theorem:4} (weighted sup-norm bound)
Let $\hat{V}^{\pi^i}$ be {an} approximate value function solution to  (\ref{eq333}), ${V}^{{\pi}^i}$ be the solution to (\ref{PICON}) and $V^{\pi^i}_P$ be the weighted sup-norm projection of ${V}^{\pi^i}$ on $\hat{\mathcal{S}}$. Then,
\begin{align}
   &  \int\limits_{\cal X}\frac{{|{{V}^{\pi^i}}-}{{{\hat{V}^{\pi^i}}\,}^{}}|}{w(x)}c(x)dx \le \frac{2}{1-{{{\Delta \gamma } }^{}}}||{{V}^{\pi^i}}-V^{\pi^i}_P\|{{}_{w }}
\label{eq46nn}
\end{align}
\end{thm}
\noindent \textit{Proof}. The proof is similar to that of Theorem \ref{Theorem:1}, and only the operator $\cal{T}$ in Theorem \ref{Theorem:1} is now replaced by the operator $\cal {\hat T}$ and thus $V^*$ is replaced by $V^{\pi}$ in the poof of Lemma \ref{Lemma:3}.  $\hfill$ $\square$ 

\begin{thm}\label{Theorem:5} (Lyapunov bound)
Let $V^{\pi^i}$ be the solution to the LEE Bellman equation (\ref{eq16}) for policy $\pi^i$, and $\hat{V}^{\pi^i}$ be the approximate solution to (\ref{eq333}). If $\hat{V}^{+}>0$ satisfies $\gamma \mathbb{M} \hat{V}^{+} < \hat{V}^{+}$, then 
\begin{align}
||V^{\pi^i}-{{\hat{V}{^{\pi^i}}}\,}||_{1,c(x)}\le \frac{2||{{\hat{V}}^{+}}|{{|}_{1,c(x)}}}{1-\Theta _{{{\hat{V}}^{+}}}^{}}\underset{\hat{V}\in \hat{\mathcal{S}}({\cal X})}{\mathop{\min }}\,||{V^{\pi^i}}\text{-}\hat{V}|{{|}_{\infty ,\frac{1}{{{\hat{V}}^{+}}}}}.   
\label{eq49nn}
\end{align}
\end{thm}
\noindent \textit{Proof.} The proof is similar to that of Theorem \ref{Theorem:2}, and is omitted due to the limited space.  $\hfill$ $\square$

Based on the model-based PI algorithm, a model-free PI algorithm is now presented utilizing the $Q$-function. Algorithm 4 uses Algorithm 3 and the relationship between $V$ and $Q$-functions, i.e., $V^{\pi}(x)=Q^{\pi}({{x},{\pi(x)}})$ to develop a model-free $Q$-based PI algorithm.







In the next subsection, details of developing a data-driven PI algorithm for the RAOC problem using sampling data are presented and provided by Algorithm 4.  


\subsection{Data-driven implementation of PI for the RAOC problem}

The main goal is to estimate the optimal risk-averse objective function $\hat{Q}^{\pi}$ in each iteration. To this aim, consider the following optimization problem. 

\noindent \textbf{Problem 7. (Approximate space $Q$-based policy evaluation) } 
\begin{align}
 &  \underset{\hat{Q}\in \hat{\mathcal{S}}({\cal X},{\cal U})}{\mathop{\max }}\,\int_{{\cal X\times \cal U}}{\hat{Q} (x,u)c(x,u)dxdu} \nonumber \\ 
 & \hat{Q}(x_i,u_i)\le l(x_i)+{{(u_{i})^{T}}R{u_{i}}}+\frac{1}{\alpha }\ln ({\mathbb{E}_{x_{u_i}}}[\exp (\alpha \gamma \hat{Q}^{}(x_{{{u_i}}}^{'},u') )]) \nonumber \\
& u'\in {\cal U} \quad i\in {\mathbb{N}_{[1,N]}}
 \label{d1}
\end{align}
 In order to incorporate the prior knowledge on the learning process,  (\ref{d1}) can be reformulated as  
\begin{align}
 &  \underset{\hat{Q}\in \hat{\mathcal{S}}({\cal X},{\cal U})}{\mathop{\max }}\,\int_{{\cal X\times \cal U}}{\hat{Q} (x,u)c(x,u)dxdu} \nonumber \\ 
& \hat{Q}(x_i,u_i)\le l(x_i)+{{(u_{i})^{T}}R{u_{i}}}+\frac{1}{\alpha }\ln ({\mathbb{E}_{x_{u_i}}}[\exp (\alpha \gamma\hat{Q}^{}(x_{{{u_i}}}^{'},\pi^k(x_i)) )]) \nonumber \\
& \quad i\in {\mathbb{N}_{[1,N]}}
 \label{d2j}
\end{align}
The difference between (\ref{d1}) and (\ref{d2j}) is that $u_i\in {\cal U}$ is fixed to be the improved control policy $\pi^k$ found from the last iteration in (\ref{d2j}) while in (\ref{d1}) $u_i\in {\cal U}$ provides certain freedom in exploring different control policies. It should be noted that (\ref{d2j}) can be regarded as a special case of (\ref{d1}). In addition, the tuple $(x_i, u_i, x_{{{u_i}}}^{'}, l(x_i))$ can be reused with adding different $u_i$. Noting the fact that (\ref{d1}) and (\ref{d2j}) contain expectation with respect to the next state of $x_i$, empirical method \cite{sutter2017} can be employed using sampling in algorithm. To this end, (\ref{d2j}) can be reformulated as 
\begin{align}
&   \underset{\hat{Q}\in \hat{\mathcal{S}}({\cal X},{\cal U})}{\mathop{\max }}\,\int_{{\cal X\times \cal U}}{\hat{Q} (x,u)c(x,u)dxdu} \nonumber \\ 
& \hat{Q}(x_i,u_i)\le l(x_i)+{{(u_{i})^{T}}R{u_{i}}}+\frac{1}{\alpha }\ln (\frac{1}{Z}\sum_{i=1}^Z{{}}[\exp (\alpha \gamma \hat{Q}^{}(x_{{{i}}}^{'},\pi^k(x_i)) )])\nonumber \\
& \quad i\in {\mathbb{N}_{[1,N]}}
 \label{d2}
\end{align}
where $x_i'$ are i.i.d. samples collected at the next state of $x_i$ given $u_i$.

Similar to \cite{banjac2019}, one effective way for guaranteeing the non-increasing risk-averse objective function is keeping all constraints from previous iterations. To alleviate the computational burden, a surrogate strategy is to only keep binding constraints from iteration $k$ for the next iteration $k+1$. The binding
constraints refer to constraints with nonzero optimal Lagrange
multipliers \cite{Boyd2004} for which their removal affects the optimal value of the convex programming. The following proposition is inspired by \cite{banjac2019}. \vspace{6pt}

\begin{prop}\label{Proposition:10}
Suppose that the risk-averse objective function in Algorithm 4 is finite for $k=1$. Then, the risk-averse value functions learned by Algorithm 4 are monotonically non-increasing.
\end{prop}
\noindent \textit{Proof.} The optimization constraints in the iteration $k+1$ includes the binding constraints $\Xi^k$ from the $k$-th iteration. Therefore, along with additional constraints, the sequence of value functions found by the optimization problem cannot increase. This completes the proof.    
$\hfill$    $\square$  


\noindent \textit{Remark 8.}  The sample trajectories of the system are assumed to be available in advance and can either come from historical data or from a generative model of the system. The value function is to be approximated with a buffer consisting of several disposal state-control-successor triples $(x_{t}^{(i)},\,u_{t}^{(i)},\,x_{t+1}^{(i)})$ where $i\in \{1,2,...,M\}$ is the experiment number. The proposed approach will allow learning an unbiased and low-variance control policy using previously collected data from the system, which is of vital importance for systems that exploration and online interaction is risky for. 


 The Data-driven $Q$-based PI algorithm is summarized in Algorithm 4. 

  \begin{algorithm}[H]
\SetAlgoLined
  \textbf{Input:}{ Parameters $\ell, \varpi,{\mathcal{Q}} \in \mathbb{R}$, basis functions ${\{q_i\}}_{i=1}^{\varpi}$, an initial policy $\pi^1$, and threshold value $\xi$ } \\
  \textbf{Set} $k=1$ and ${{\Xi }^{0}}=\emptyset$
  
\While{ $\null \left| {{{\hat Q}^i} - {{\hat Q}^{i - 1}}} \right| \ge \xi $}{
  \textbf{\textup{for}} $q=1,..., {\mathcal{Q}}$ \textup{\textbf{do}} \\
    $\null$    $\quad$  Collect $x_q \in {\cal X}$, apply $u_q \in {\cal U}$, observe the next state of $x_q$, $x_{u_q}^{'}$, and $l_q \in \mathbb{R}_{>0}$   \\
  \textup{\textbf{end}} \\ 
  $\textbf{\textup{for}}$ $i=1,..., \ell  $ \textup{\textbf{do}}\\ 
    $\null$ $\quad$   Select $(x_i, u_i, x_{u_i}^{'}, l_i)$ $\in {\{(x_q,  u_q, x_{u_q}^{'},  l_q)}\}_{q=1}^{\mathcal{Q}}$  \\
    $\null$ $\quad$   Select $u_i^{'}={\pi}^{k(x_i)}$
\begin{align}
&   {{\hat{Q}}^{{{\pi }^{i}}}}\leftarrow \underset{\hat{Q}^i\in \hat{\mathcal{S}}({\cal X},{\cal U})}{\mathop{\max }}\,\int_{{\cal X\times \cal U}}{\hat{Q}^{i} (x,u)c(x,u)dxdu} \nonumber \\ 
&  \null \quad \quad \quad \hat{Q}^i(x_i,u_i)\le l(x_i)+{{(u_i)^{T}}R{u_i^{}}} \nonumber \\
& \quad \quad \quad  +\frac{1}{\alpha }\ln (\frac{1}{Z}\sum_{i=1}^Z{{}}[\exp (\alpha \gamma \hat{Q}^i(x_{{{u_i}}}^{'},u_i') )]) \nonumber \\
&  \quad \quad \forall (x_i,u_i,u_i') \in {\cal X} \times {\cal U} \times {\cal U},     i \in {\mathbb{N}}_{[ 1, \ell]} \nonumber \\
&  \null \quad \quad \quad \hat{Q}^i(x_j,u_j)\le
  l(x_j)+{{(u_j)^{T}}R{u_j^{}}} \nonumber \\
& \quad \quad \quad +\frac{1}{\alpha }\ln (\frac{1}{Z}\sum_{i=1}^Z{{}}[\exp (\alpha \gamma \hat{Q}^i(x_{{{u_j}}}^{'},u_j') )]) \nonumber \\
&   \quad \quad \forall (x_j,u_j,u_j') \in {\cal X} \times {\cal U} \times {\cal U},  j \in  \Xi^{k-1}
\label{eq39}
\end{align}
  $\quad$ $\Xi^{k}\leftarrow \textup{Binding constraints}$ \\
  $\null$ $\quad \pi^{k+1}(x)\leftarrow \mathop {\arg \min }\limits_{u \in {\cal U}} {\hat Q^{{\pi ^i}}}(x,u), \quad \forall x\in {\cal X}$\\
  $\null$ $\quad$ $k \leftarrow k+1$ \\
  \textup{\textbf{end}}
}
 \caption{Data-driven $Q$-based PI algorithm.}
\end{algorithm}

\noindent \textit{Remark 9.} The summarized Algorithm 4 can be regarded as an actor-critic structure for optimal risk-averse value function approximation. The critic estimates the $\hat{Q}^{\pi^i}$ in each iteration which is followed by a greedy actor to improve control policy. Since the function space is restricted on (\ref{eq29}) so the improved control policy can be calculated efficiently. Notice that the number of constraints in (\ref{eq39}) is bounded by a finite number. $Q$ data tuples drawn from the experiment and the number of derived constraints is $N$. To accelerate the convergence of the algorithm without extra computational burden, the binding constraints from the last iteration $k$ are included as additional constraints in the LP solved in iteration $k+1$. The total number of the binding constraints is assumed to be bounded.

Note that for the policy evaluation step, while samples are collected from the current policy to form the optimization constraints, the optimization can still use previous samples, and, if the sample set is large enough, it will also include samples from the current policy.

\section{Numerical Results}

\begin{figure}[!t]
\centering{\includegraphics [width=3.5in] {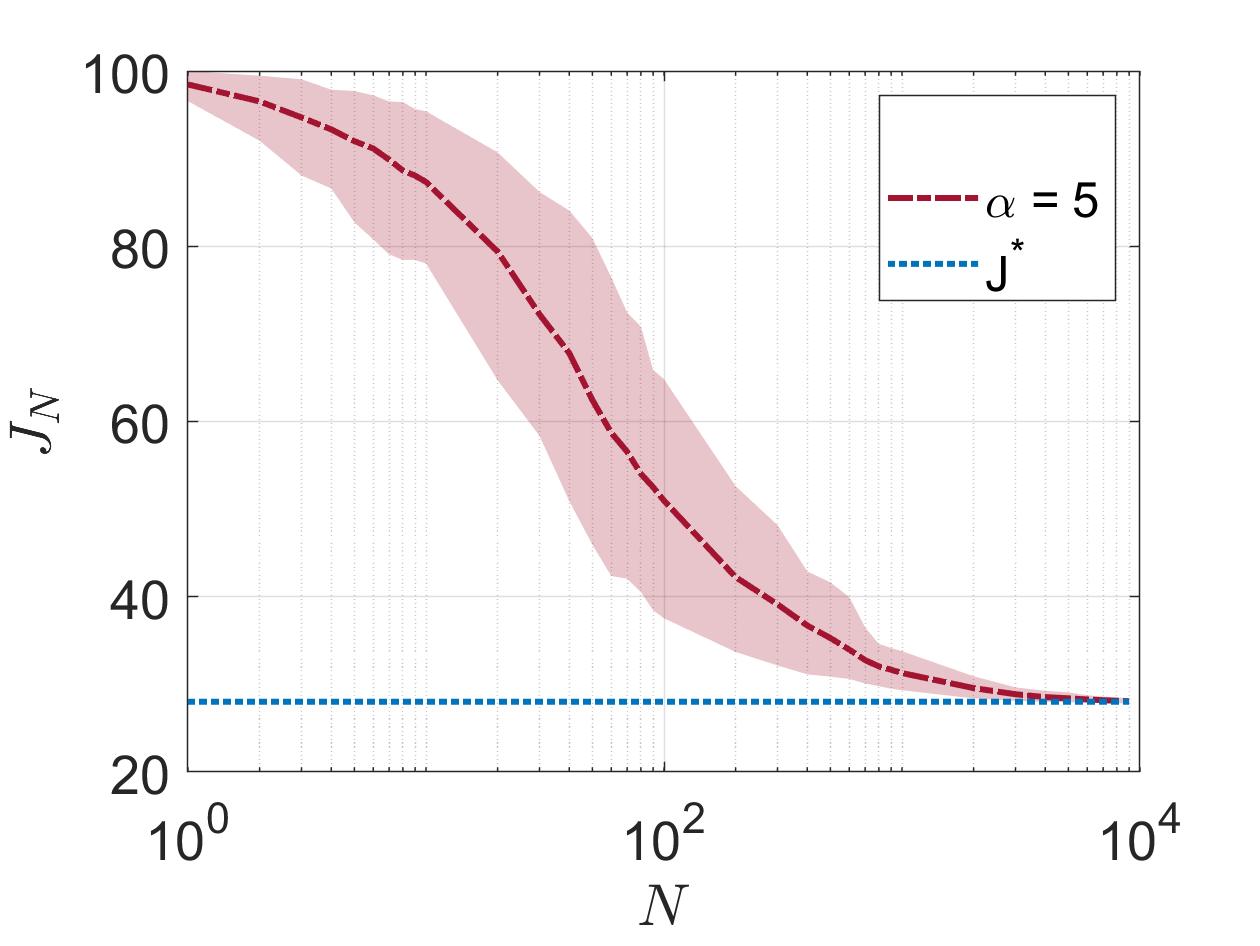}}
\caption{The approximated performance function $J_{N}$ with $\alpha = 5$ versus the number of sampled constraints ${N}$.} \label{fig3a5}
\end{figure}

\begin{figure}[!t]
\centering{\includegraphics [width=3.5in] {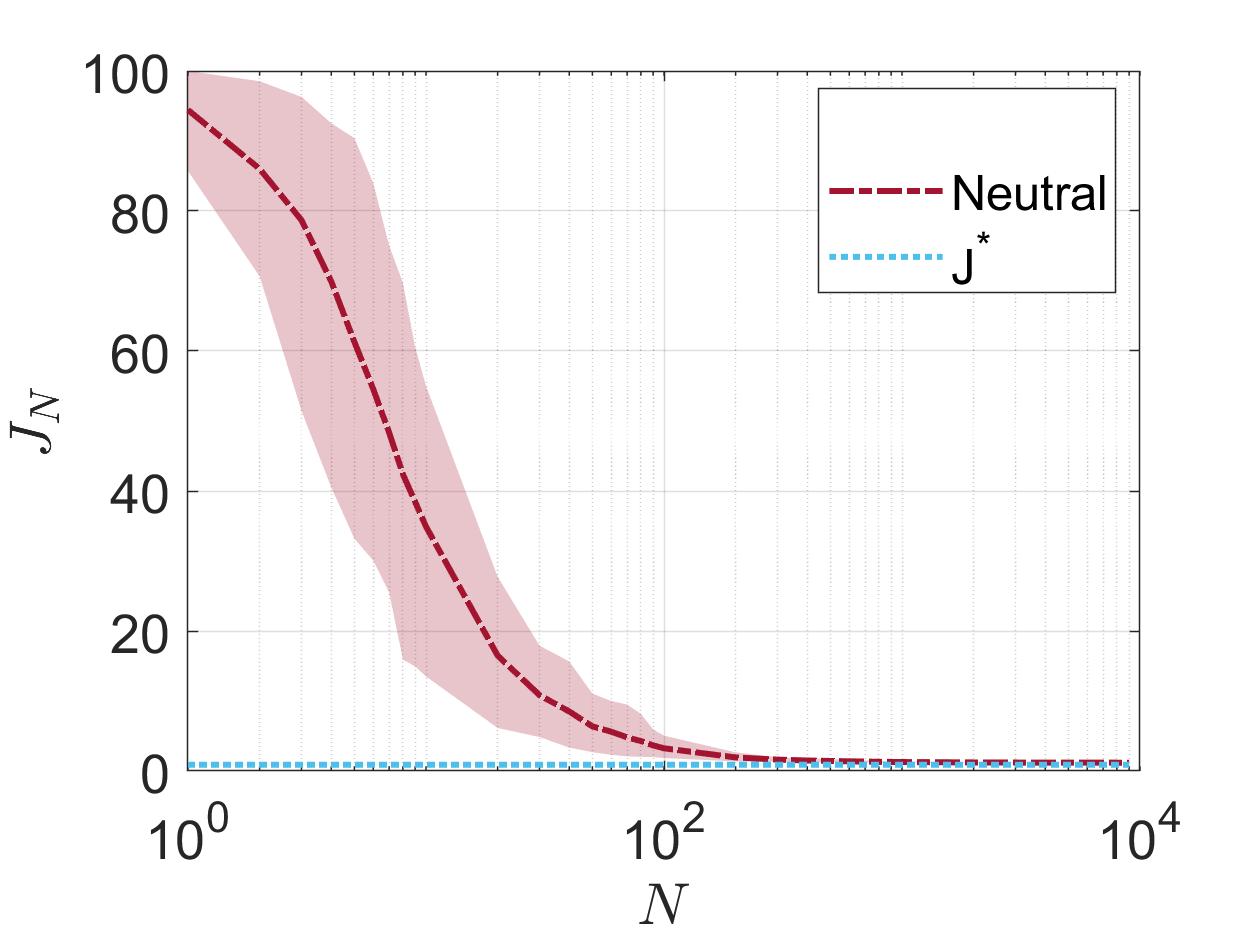}}
\caption{The approximated performance function $J_{N}$ for the risk neutral case versus the number of sampled constraints ${N}$.} \label{fig2N}
\end{figure}

\begin{figure}[!t]
\centering{\includegraphics [width=3.5in] {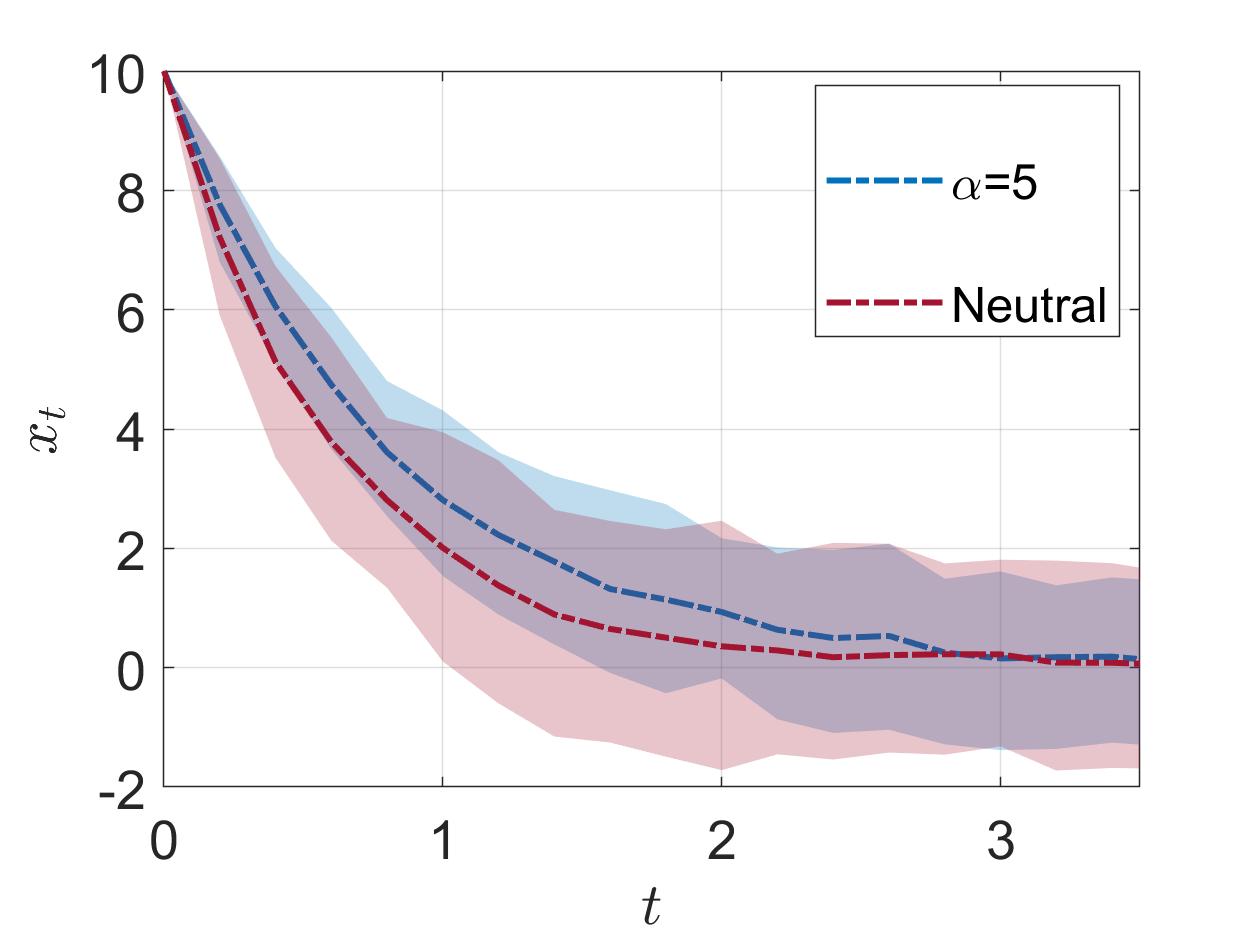}}
\caption{System state trajectories for the risk-averse case with $\alpha = 5$ and the risk-neutral case.} \label{fig4}
\end{figure}

The efficiency of the proposed algorithm is verified using a one dimensional LQG numerical example from \cite{sutter2017} with linear dynamics given by
\begin{align}
{x_{t + 1}} = 0.8{x_t} + 0.5{u_t} + {\varepsilon _t}
\label{ex1}
\end{align}
where ${\cal X} = [ - 10,10]$ and ${\cal U} = [ - 10,10]$. The risk-neutral cost function is assumed to be  ${{\cal L}^\pi } = \sum\limits_{t = 0}^\infty  {{\gamma ^t}({x^2} + 0.5{u^2})}$, with $\gamma  = 0.95$. The disturbances  ${\left\{ {{\varepsilon _t}} \right\}_{t \in \mathbb{N}}}$  are i.i.d. random variables generated by a truncated normal distribution with known parameters $\mu  = 0$  and $\sigma  = 1$. That is, the process ${\varepsilon _t}$  has a distribution density given by
\begin{align}
{D_{{\varepsilon _t}}}(x,\mu  = 0,\sigma  = 1) = \left\{ {\begin{array}{*{20}{c}}
{\frac{{\phi \left( x \right)}}{{\Phi \left( {10} \right) - \Phi \left( { - 10} \right)}},}&{x \in {\cal X}}\\
0&{{\rm{ o}}{\rm{.w}}{\rm{. }}}
\end{array}} \right. \label{eq:ex21}
\end{align}
where $\phi (.)$  and $\Phi (.)$ are, respectively, probability density functions of the standard normal distribution and its cumulative distribution function. To approximate value functions, Fourier basis functions are used in the simulation as $\left\{ {\frac{{10}}{\pi }\cos \left( {\frac{{\pi s}}{{10}}} \right),\frac{5}{\pi }\sin \left( {\frac{{\pi s}}{5}} \right),\frac{{10}}{{3\pi }}\cos \left( {\frac{{3\pi s}}{{10}}} \right),\frac{5}{{2\pi }}\sin \left( {\frac{{2\pi s}}{5}} \right)} \right\}$. The uniform distribution on ${\cal X} \times {\cal U} = [ - 10,10] \times [ - 10,10]$  is used to generate the random samples.

Fig.~\ref{fig3a5} shows the actual performance as a function of number of samples for $\alpha=5$. One can observe that using at least $9e^4$ sampled constraints the proposed approach can reach the optimal performance function ${J^*}$ with an acceptable approximation accuracy. Note that the blue dotted line in Fig.~\ref{fig3a5} is the optimal solution ${J^*}$ approximated by $n=10$ (number of basis functions) and $N=1e^7$ (number of sampled constraints). Moreover, the shaded area and dashed line represent the results between $[10 \%, 90 \%]$ quantiles and the means across $100$ independent experiments. One can see from Fig.~\ref{fig3a5} that by choosing a greater number for sampled constraints the variance of the approximation will be diminished significantly. Fig.~\ref{fig2N} shows the performance of Algorithm 1 for the risk-neutral case (i.e., $\alpha =0$), in which we aim to minimize the expectation of the quadratic cost function (\ref{eq2}), i.e., $J= {\mathbb{E} }[\cal{L}^{\pi}]$. Note that the blue dotted line in Fig.~\ref{fig2N} is the optimal solution ${J^*}$ approximated by $n=10$ (number of basis functions) and $N=1e^4$ (number of sampled constraints). Moreover, the shaded area and dashed line represent the results between $[10 \%, 90 \%]$ quantiles and the means across $100$ independent experiments. The trajectories of state for both risk-neutral and risk-averse cases are displayed in Fig.~\ref{fig4}. Note that the variance of the performance is lower for the risk-averse case as expected. If fact, in the risk-averse case (i.e., $\alpha > 0$), control acts as if the process ${\varepsilon _t}$ directed the state in an undesired direction, and therefore it tries to reduce the effects of the process ${\varepsilon _t}$ on state $x_t$ through minimizing the expected value and the variance of the cost \eqref{eq2}.




\section{Conclusion}
In this paper, three model-free reinforcement learning (RL) algorithms are developed to solve the risk-averse optimal control (RAOC) problem for discrete-time nonlinear systems. The first algorithm turns the learning process of finding a control policy into a one-shot convex optimization problem using only a large dataset of a priori collected samples. For the cases where online interaction is also necessary because rich data samples are not available a priori, iterative VI and PI-based RL algorithms are presented to learn the solution to the RAOC problem. The advantage of the VI algorithm is that it solves an LP optimization at each iteration and the advantage of the PI algorithm is that it guarantees stability of learned control policies. Convergence of the presented data-based algorithms to the optimal solution is shown for the exact convex program-based algorithms, which are infinite dimensional, and the approximate error bounds due to function approximation and constraint sampling for the resulting standard convex  optimization algorithms are found. A simulation example is provided to validate the theoretical results. The future work is to provide sample complexity analysis for data-based algorithms and for the stability guarantee of the PI algorithm.


\appendices

  \section{proof of Lemma \ref{Lemma:1}}
  \label{proof of exp and var }
It is well-known that $\exp (f)$ and $\ln (1+f)$, respectively, have Maclaurin series $\exp (f)=1+f+{{f}^{2}}/2+ {\cal O}({{f}^{3}})$ and $\ln (1+f)=f-{{f}^{2}}/2+ {\cal O}({{f}^{3}})$ which converge for all $f$. Using these facts, substitute the first series into the second one, easily omit terms of higher order than the $3$-th degree denoted by using the big ${\cal O}$ notation, and some manipulation, then, one has 
\begin{align}
&\ln ({\mathbb{E}_{{x_{\pi}'} }}(\exp (\alpha {{\mathcal{L}}^{\pi }}(.))))    \approx   \alpha {\mathbb{E}_{{x_{\pi}'} }}({{\mathcal{L}}^{\pi }}(.))+ \nonumber \\ & \left( {{\alpha }^{2}}/2 \right){\mathbb{E}_{{x_{\pi}'} }}({{({{\mathcal{L}}^{\pi }}(.))}^{2}})
-(1/2){{\left( \alpha {\mathbb{E}_{{x_{\pi}'} }}({{\mathcal{L}}^{\pi }}(.))+\left( {{\alpha }^{2}}/2 \right) \right)}^{2}} \label{majid1}  
\end{align}  
Expanding the squares, dividing both sides of (\ref{majid1}) by $\alpha $, and omitting $O(\alpha^3)$ based on the same reasoning given in \cite{Saldi2020}, we have
\begin{align}
&     \frac{1}{\alpha }\ln ({\mathbb{E}_{{x_{\pi}'} }}(\exp (\alpha {{\mathcal{L}}^{\pi }}(.)))) \approx \nonumber \\ &{\mathbb{E}_{{x_{\pi}'} }}({{\mathcal{L}}^{\pi }}(.)) 
 +\left( \alpha /2 \right)\left( {\mathbb{E}_{{x_{\pi}'} }}({{({{\mathcal{L}}^{\pi }}(.))}^{2}})-{{({\mathbb{E}_{{x_{\pi}'} }}({{\mathcal{L}}^{\pi }}(.)))}^{2}} \right)  \label{majid2}
\end{align}   
which completes the proof.  $\hfill$ $\square$ 

 \section{proof of Theorem \ref{Proposition:2}}
  
The proof is shown only for $\cal T$ operator as the proof for $\hat {\cal} T$ is similar. Before proceeding with the proof, note that using the definition of weighted sup-norm, one has
 \begin{align}
  & \alpha \gamma {V^{\pi_2}}(x)-\alpha \gamma {V^{\pi_1}}(x)\le \alpha \gamma ||{V^{\pi_1}}-{V^{\pi_2}}|{{|}_{w}}w(x) \nonumber \\ 
 & \Leftrightarrow \,\,\alpha \gamma {V^{\pi_2}}(x)-\alpha \gamma ||{V^{\pi_1}}-{V^{\pi_2}}|{{|}_{w}}w(x)\le \alpha \gamma {V^{\pi_1}}(x) \nonumber \\ 
 & \Leftrightarrow \,\ln {\mathbb{E}_{x}}[\exp (\alpha \gamma {V^{\pi_2}}(x)-\alpha \gamma ||{V^{\pi_1}}-{V^{\pi_2}}|{{|}_{w}}w(x))]\, \nonumber \\ 
 & \,\,\,\,\,\,\,\le \ln {\mathbb{E}_{x}}[\alpha \gamma {V^{\pi_1}}(x)] \nonumber \\ 
 & \Leftrightarrow -\frac{1}{\alpha }\,\ln {\mathbb{E}_{x}}[\exp (\alpha \gamma {V^{\pi_2}}(x)-\alpha \gamma ||{V^{\pi_1}}-{V^{\pi_2}}|{{|}_{w}}w(x))] \nonumber \\ 
 & \,\,\,\,\,\,\,\ge -\frac{1}{\alpha }\ln {{\mathbb{E}}_{x}}[\alpha \gamma {V^{\pi_1}}(x)] w(x)
 \label{mono}
\end{align}
Then,
\begin{align}
 &  {\cal T}{V^{\pi_2}(x)}-{\cal T}{V^{\pi_1}(x)} \le \nonumber \\
 & \underset{u\in {\cal U}}{\mathop{\sup }}\,\big ( -\frac{1}{\alpha }\ln {{\mathbb{E}}_{x'_u}}[\exp \alpha \gamma {V^{\pi_1}}(x'_u)]+\frac{1}{\alpha }\ln {{\mathbb{E}}_{x'_u}}[\exp \alpha \gamma {V^{\pi_2}}(x'_u)] \big ) \nonumber  \\ 
  & \le \underset{u\in {\cal U}}{\mathop{\sup }}\,\big ( -\frac{1}{\alpha }\ln {{\mathbb{E}}_{x'_u}}[\exp (\alpha \gamma ||{V^{\pi_1}}-{V^{\pi_2}}|{{|}_{w}}w(x'_u)+\alpha \gamma {V^{\pi_2}}(x'_u))] \nonumber  \\
 &  + \frac{1}{\alpha }\ln {{\mathbb{E}_{{x'_u}}}[\exp \alpha \gamma {V^{\pi_2}(x'_u)}] \big) }    \nonumber  \\ 
&  \le \mathop {\sup }\limits_{u \in {\cal U}} {\mkern 1mu} \big( - \frac{1}{\alpha }\ln {\rm{ }}{\mathbb{E}}_{x'_u}[\exp (\alpha \gamma ||V^{\pi_1} - V^{\pi_2}|{|_w}w({x'_u})] \nonumber  \\
&  -\frac{1}{\alpha }\ln {{\mathbb{E}}_{x'_{u}}}[\exp \alpha \gamma {V^{\pi_2}}(x'_{u})]+\frac{1}{\alpha }\ln {{\mathbb{E}}_{x'_{u}}}[\exp \alpha \gamma {V^{\pi_2}}(x'_{u})] \big ) \nonumber \\
 &   =\underset{u\in {\cal U}}{\mathop{\sup }}\,\big ( -\frac{1}{\alpha }\ln {{\mathbb{E}}_{x'_{u}}}[\exp (\alpha \gamma ||{V^{\pi_1}}-{V^{\pi_2}}|{{|}_{w}}w(x'_{u})] \big ) \nonumber
 \end{align}
\begin{align} 
 &  \underset{u\in {\cal U}}{\mathop{\le \sup }}\,\big ( -\frac{1}{\alpha }{{\mathbb{E}}_{x'_{u}}}\ln [\exp (\alpha \gamma ||{V^{\pi_1}}-{V^{\pi_2}}|{{|}_{w}}w(x'_{u})] \big ) \nonumber  \\ 
 &  \underset{u\in {\cal U}}{\mathop{=\sup }}\,\big ( \frac{1}{\alpha }{{\mathbb{E}}_{x'_{u}}}(\alpha \gamma ||{V^{\pi_1}}-{V^{\pi_2}}|{{|}_{w}}w(x'_{u})) \big ) \nonumber\\ 
 &  \underset{u\in {\cal U}}{\mathop{\,\,=\sup }}\,\big ( \gamma {{\mathbb{E}}_{x'_{u}}}(||{V^{\pi_1}}-{V^{\pi_2}}|{{|}_{w}}w(x'_{u})) \big ) \nonumber \\ 
 &  \le \Delta \gamma ||{V^{\pi_1}}-{V^{\pi_2}}|{{|}_{w}}w(x'_u)\, 
\end{align}
where the second inequality follows from (\ref{mono}), the third inequality from $\ln\mathbb{E}[f(x)+g(x)]\ge \mathbb{E}(f(x))+\mathbb{E}(g(x))$, the fourth inequality from from Jensen's inequality  and the last one from Assumption 1.  By changing the roles of ${V^{\pi_1}}$ with ${V^{\pi_2}}$ we obtain \eqref{cont1}. This completes the proof. $\hfill$ $\square$

  \section{proof of Lemma \ref{Lemma:3}}
  \label{proof of convexity of }
Corollary \ref{Corollary:1} showed that any feasible solution of the optimization Problem 3 satisfies the optimal LEE Bellman operator $\hat V \le \cal{T} \hat V$. Based on Propositions \ref{Proposition:1} and \ref{Proposition:2}, the optimal LEE Bellman operator is monotone and contracting with a fixed-point solution $V^{*}$. Therefore, for any $\hat{V}\le {\cal T}\hat{V}$, one has
\begin{align}
\hat{V}\le {\cal T}\hat{V}\le {\cal T}^2\hat{V}\le ...\le V^{*}
 \label{lemma2}
\end{align}
It follows from (\ref{lemma2}) that
\begin{align}
   ||{{V}^{*}}-\hat{V}||_{1,c(x)} & = \int_{\cal{X}}|V^{*}-\hat{V}(x)|c(x) dx  \nonumber \\  
   & = V^{*}-\int_{\cal{X}}\hat{V}(x)c(x) dx 
 \label{eqlemma2}
\end{align}
Therefore, maximizing the $\int_{X}\hat{V}(x)c(x) dx$ (which is (\ref{eq30n})) equals to minimizing  $||{{V}^{*}}-\hat{V}||_{1,c(x)}$. This completes the proof. $\hfill$ $\square$

   \section{proof of Theorem \ref{Theorem:1}}
  \label{proof of Theorem 1}

Based on Proposition \ref{Proposition:2}, the operator $\hat{\cal T}$ is contraction map, and therefore, the following inequality holds
\begin{align}
||{{V}^{*}}-{\hat{\cal T}^{}}V^P|{{|}_{w }}=||\hat{{\cal T}^{}}{{V}^{*}}-{\hat{\cal T}^{}}V^P|{{|}_{w }}\le {{\Delta \gamma }^{}}||{{V}^{*}}-V^P|{{|}_{w }}
\label{eq42}
\end{align}
Based on (\ref{eq42}), the following inequalities hold
\begin{align}
{{\Delta \gamma }^{}}||{{V}^{*}}-V^P|{{|}_{w }}\ge \frac{{{V}^{*}}-{{\cal T}^{}}V^P}{w(x)}\ge -{{\Delta \gamma }^{}}||{{V}^{*}}-{{\cal T}^{}}V^P|{{|}_{w }}
\label{eq43}
\end{align}
which results in
\begin{align}
  & \frac{{\hat{\cal T}^{}}V^P}{w(x)}\ge \frac{{{V}^{*}}}{w(x)}-{{{\Delta \gamma } }^{}}||{{V}^{*}}-V^P|{{|}_{w }} 
  \ge \frac{V^P}{w(x)}-||{{V}^{*}}-V^P|{{|}_{w }} \nonumber \\
 & -{{{\Delta \gamma } }^{}}||{{V}^{*}}-V^P|{{|}_{w }} =\frac{V^P}{w(x)}-(1+{{{\Delta \gamma } }^{}})||{{V}^{*}}-V^P|{{|}_{w }} 
\label{eq44}
\end{align}
Then, define $ \hat{V}\,\in \mathfrak{S}$ as
\begin{align}
   \frac{\hat{V}}{w(x)}=\frac{V^P}{w(x)}-\frac{1+{{{\Delta \gamma } }^{}}}{1-{{{\Delta \gamma } }^{}}}||{{V}^{*}}-V^P|{{|}_{w}}
\label{eq45}
\end{align}
Then, $\hat{V}$ also satisfies $\hat{V}\le {\cal T}\hat{V}$ since the second term of right-hand side is a downwards shift term and the approximate space $\mathcal{S}$ allows for affine combinations of basis functions. Based on (\ref{eq44}) and (\ref{eq45}), the following holds
\begin{align}
 &  \int\limits_{\cal X}\frac{{|{{V}^{*}}-}{{{\hat{V}}\,}^{*}}|}{w(x)}c(x)dx\le \int\limits_{{\cal X}}\frac{{|{{V}^{*}}-}\hat{V}|}{w(x)}c(x)dx  \nonumber \\ 
&  \le ||{{V}^{*}}-\hat{V}|{{|}_{w }}\le ||{{V}^{*}}-V^P\,|{{|}_{w }}+||V^P\,-\hat{V}|{{|}_{w }} \nonumber \\ 
&  =||{{V}^{*}}-V^P\,|{{|}_{w }}+\frac{1+{{{\Delta \gamma } }^{}}}{1-{{{\Delta \gamma }}^{}}}||{{V}^{*}}-V^P|{{|}_{\infty }} \nonumber \\ 
& =\frac{2}{1-{{{\Delta \gamma } }^{}}}||{{V}^{*}}-V^P\|{{|}_{w }}
\label{eq46}
\end{align}
where ${{\hat{V}}\,}^{*}$ is the solution to (\ref{eq30n}). This completes the proof.

  \section{proof of Lemma \ref{Lemma:4}}
  \label{proof of the Lemma 4}
  For any given function $g_1(.)$ and $g_2(.)$, one has
\begin{align}
 & {\cal T} {{g}_{1}(x)}- {\cal T} {{g}_{2}(x)} = \nonumber \\ 
 &  \underset{u}{\mathop{\min }}\,\{{{l}_{{{g}_{1}}}}(x)+\frac{u_{{{g}_{1}}}^{T}R{{u}_{{{g}_{1}}}}}{2}+\frac{1}{\alpha }\ln (\mathbb{E}(\exp(\alpha \gamma {{g}_{1}}(f(x,u,\varepsilon ))))\} \nonumber \\ 
 & -\underset{u}{\mathop{\min }}\,\{{{l}_{{{g}_{2}}}}(x)+\frac{u_{{{g}_{2}}}^{T}R{{u}_{{{g}_{2}}}}}{2}+\frac{1}{\alpha }\ln (\mathbb{E}(\exp(\alpha \gamma {{g}_{2}}(f(x,u,\varepsilon ))))\} \nonumber \\ 
&  ={{l}_{{{g}_{1}}}}(x)+\frac{\overset{\text{*}}{\mathop{u}}\,_{{{g}_{1}}}^{T}R{{\overset{\text{*}}{\mathop{u}}\,}_{{{g}_{1}}}}}{2}+\frac{1}{\alpha }\ln (\mathbb{E}(\exp(\alpha \gamma {{g}_{1}}(f(x,{{\overset{\text{*}}{\mathop{u}}\,}_{{{g}_{1}}}},\varepsilon )))) \nonumber \\ &  \,\,\,-{{l}_{{{g}_{2}}}}(x)-\frac{\overset{\text{*}}{\mathop{u}}\,_{{{g}_{2}}}^{T}R{{\overset{\text{*}}{\mathop{u}}\,}_{{{g}_{2}}}}}{2}-\frac{1}{\alpha }\ln (\mathbb{E}(\exp(\alpha \gamma {{g}_{2}}(f(x,{{\overset{\text{*}}{\mathop{u}}\,}_{{{g}_{2}}}},\varepsilon )))) \nonumber \\ 
&  \ge  {{l}_{{{g}_{1}}}}(x)+\frac{\overset{\text{*}}{\mathop{u}}\,_{{{g}_{1}}}^{T}R{{\overset{\text{*}}{\mathop{u}}\,}_{{{g}_{1}}}}}{2}+\frac{1}{\alpha }\ln (\mathbb{E}(\exp(\alpha \gamma {{g}_{1}}(f(x,{{\overset{\text{*}}{\mathop{u}}\,}_{{{g}_{1}}}},\varepsilon )))) \nonumber  \\
&  \,\,-{{l}_{{{g}_{1}}}}(x)-\frac{\overset{\text{*}}{\mathop{u}}\,_{{{g}_{1}}}^{T}R{{\overset{\text{*}}{\mathop{u}}\,}_{{{g}_{1}}}}}{2}-\frac{1}{\alpha }\ln (\mathbb{E} (\exp(\alpha \gamma {{g}_{2}}(f(x,{{\overset{\text{*}}{\mathop{u}}\,}_{{{g}_{1}}}},\varepsilon ))))  \nonumber \\
&   =\frac{1}{\alpha }\ln (\mathbb{E} (\exp(\alpha \gamma {{g}_{1}}(f(x,{{\overset{\text{*}}{\mathop{u}}\,}_{{{g}_{1}}}},\varepsilon )))) \nonumber\\
&\,\, -\frac{1}{\alpha }\ln (\mathbb{E} (\exp(\alpha \gamma {{g}_{2}}(f(x,{{\overset{\text{*}}{\mathop{u}}\,}_{{{g}_{1}}}},\varepsilon ))))\nonumber \\
&   = \frac{1}{\alpha }\ln (\frac{\mathbb{E} (\exp(\alpha \gamma {{g}_{1}}(f(x,{{\overset{\text{*}}{\mathop{u}}\,}_{{{g}_{1}}}},\varepsilon )))}{\mathbb{E}(\exp(\alpha \gamma {{g}_{2}}(f(x,{{\overset{\text{*}}{\mathop{u}}\,}_{{{g}_{1}}}},\varepsilon )))}) \nonumber \\
&   =\frac{1}{\alpha }\ln (\mathbb{E}(\exp (\alpha \gamma ({{g}_{1}}(f(x,{{\overset{\text{*}}{\mathop{u}}\,}_{{{g}_{1}}}},\varepsilon))-{{g}_{2}}({{f}_{}}(x,{{\overset{\text{*}}{\mathop{u}}\,}_{{{g}_{1}}}},\varepsilon)))))) \nonumber\\
& \ge \gamma \mathbb{E} (g_1({{f}_{}}(x,{{\overset{\text{*}}{\mathop{u}}\,}_{{{g}_{2}}}},\varepsilon))-g_2({{f}_{}}(x,{{\overset{\text{*}}{\mathop{u}}\,}_{{{g}_{2}}}},\varepsilon)))
  \label{eq61}
\end{align}
where ${{\overset{\text{*}}{\mathop{u}}\,}_{{{g}_{1}}}}$ and ${{\overset{\text{*}}{\mathop{u}}\,}_{{{g}_{2}}}}$ denote the greedy policy with respect to $g_1$ and $g_2$, respectively. The Last inequality follows the Jensen's inequality. Therefore
\begin{align}
 & {\cal T} {{g}_{2}(x)}- {\cal T} {{g}_{1}(x)} \le \gamma \mathbb{E} (g_2({{f}_{1}}(x,{{\overset{\text{*}}{\mathop{u}}\,}_{{{g}_{2}}}},\varepsilon))-g_1({{f}_{1}}(x,{{\overset{\text{*}}{\mathop{u}}\,}_{{{g}_{2}}}},\varepsilon))) \nonumber\\
 & \,\,\,\,\,\,\,\,\,\quad \quad \le \gamma \underset{u}{\mathop{\max }} \mathbb{E} |g_2({{f}_{1}}(x,{{\overset{\text{*}}{\mathop{u}}\,}_{{{g}_{2}}}},\varepsilon))-g_1({{f}_{1}}(x,{{\overset{\text{*}}{\mathop{u}}\,}_{{{g}_{2}}}},\varepsilon))|\nonumber\\
 & \,\,\,\,\,\,\,\,\,\quad \quad = \gamma \mathbb{M} (g_2({}x)-g_1({}x))
 \label{lemma4}
\end{align}

This completes the proof. $\hfill$ $\square$ 
  
  \section{proof of Lemma \ref{Lemma:5}}
  \label{ThirdAppendix}
For any $J(.) >0$ and any $V(.)$, the following holds
\begin{align}
|V^*(x)-V(x)|\le ||V^*(x)-V(x)||_{\infty, \frac{1}{J}}J(x) 
\label{eq64}
\end{align}
Based on (\ref{lemma4}), 
\begin{align}
 |{\cal T}V^*(x)-{\cal T}V(x)| & \le \gamma \mathbb{M}(|V^*(x)-V(x)|) \nonumber \\ 
& \le \gamma ||V^*-V||_{\infty, \frac{1}{J}}\mathbb{M}(J(x))
\label{eq65}
\end{align}
Defining $\delta=||V^*-V||_{\infty, \frac{1}{J}}$, then \begin{align}
   ({\cal T} V)(x) & \ge {{V}^{*}}(x)-\gamma \delta  \mathbb{M}(V(x)) \nonumber \\ 
  & \ge V(x)-\delta  J(x)-\gamma \delta  \mathbb{M}(J(x)) 
\label{eq66}
\end{align}
This completes the proof. $\hfill$ $\square$

    \section{proof of  Lemma \ref{Lemma:6}}
  \label{FourthAppendix}
  
It follows from (\ref{eq:44lema}) that 
\begin{align}
& \text{ }\!\!|\!\!\text{ }{\cal T} V(x)- { \cal T} \overline{V}(x)| = \nonumber \\ 
&  |{\cal T} V(x)-{\cal T} \big(V-||{{V}^{*}}-V|{{|}_{\infty ,\frac{1}{V_L}}}(\frac{2}{1-{{\Theta }_{V_L}}}-1)V_L(x)\big)| \nonumber \\ 
&  \le \gamma \underset{u\in {\cal U}}{\mathop{\max }}\, \mathbb{E}(||{{V}^{*}}-V|{{|}_{\infty ,\frac{1}{V_L}}}(\frac{2}{1-{{\Theta }_{V_L}}}-1)V_L(f(x,u,\varepsilon ))) \nonumber \\ 
&  =\gamma ||{{V}^{*}}-V|{{|}_{\infty ,\frac{1}{V_L}}}(\frac{2}{1-{{\Theta }_{V_L}}}-1)\mathbb{M} (V_L(x))  
\label{eq67}
\end{align}
Using (\ref{eq66}) and (\ref{eq67}), the following inequality holds
\begin{align}
 & {\cal T} \overline{V}(x)\ge V-||{{V}^{*}}-V|{{|}_{\infty ,\frac{1}{{{V}_{L}}}}}(\gamma \mathbb{M} ({{V}_{L}})+{{V}_{L}}) \nonumber \\ 
 & \,\,\,\,\,\,\,\,\,\,\,\,\,\,\,-\gamma ||{{V}^{*}}-V|{{|}_{\infty ,\frac{1}{{{V}_{L}}}}}(\frac{2}{1-{{\Theta }_{{{V}_{L}}}}}-1)\mathbb{M} ({{V}_{L}}) \nonumber \\ 
 & \,\,\,\,\,\,\,\,\,\,\,\,\,\,\,=\overline{V}-||{{V}^{*}}-V|{{|}_{\infty ,\frac{1}{{{V}_{L}}}}}(\gamma \mathbb{M} ({{V}_{L}})+{{V}_{L}}) \nonumber \\ 
 & \,\,\,\,\,\,\,\,\,\,\,\,\,\,\,-||{{V}^{*}}-V|{{|}_{\infty ,\frac{1}{{{V}_{L}}}}}(\frac{2}{1-{{\Theta }_{{{V}_{L}}}}}-1)(\gamma \mathbb{M} ({{V}_{L}})-{{V}_{L}})  \label{eq68}
\end{align}
Since $V_L$ is the Lyapunov function, the following equality holds
\begin{align}
 \frac{2}{1-{{\Theta }_{{{V}_{L}}}}}-1 & =\frac{2}{1-{\mathop {\max }\limits_{x \in {\cal X}} }(\lambda \mathbb{M} ({{V}_{L}}(x)))/{{V}_{L}}(x)}-1 \nonumber \\ 
 & ={\mathop {\max }\limits_{x \in {\cal X}}}\,\frac{{{V}_{L}}(x)+\delta \mathbb{M} ({{V}_{L}}(x))}{{{V}_{L}}(x)-\delta \mathbb{M} ({{V}_{L}}(x))} 
    \label{eq69}
\end{align}
Plugging (\ref{eq69}) into (\ref{eq68}) yields 
\begin{align}
{\cal T}\overline{V}(x) 
& \ge \overline{V}-||{{V}^{*}}-V|{{|}_{\infty ,\frac{1}{{{V}_{L}}}}}(\gamma \mathbb{M} ({{V}_{L}})+{{V}_{L}})
\nonumber \\ 
& +||{{V}^{*}}-V|{{|}_{\infty ,\frac{1}{{{V}_{L}}}}}(\gamma \mathbb{M} ({{V}_{L}})+{{V}_{L}}) =\overline{V}(x) 
 \label{eq70}
\end{align}
This completes the proof. $\hfill$ $\square$

    \section{proof of Theorem \ref{Theorem:2}}
  \label{proof of Theorem 2}
\begin{align}
&   ||{{V}^{*}}-{{\hat{V}}^{*}}|{{|}_{1,c(x)}}\le |\text{ }\!\!|\!\!\text{ }{{V}^{*}}-\overline{V}|{{|}_{1,c(x)}}     \nonumber \\ 
& \qquad =\int\limits_{\cal X}{c(x){{V}_{L}}(x)\frac{|{{V}^{*}}(x)-\overline{V}|}{{{V}_{L}}(x)}} \nonumber \\ 
& \qquad \le \int\limits_{\cal X}{c(x){{V}_{L}}(x) {\mathop {\max }\limits_{x \in {\cal X}} }}\frac{|{{V}^{*}}(x)-\overline{V}|}{{{V}_{L}}(x)} \nonumber  \\ 
& \qquad =||{{V}_{L}}|{{|}_{1,c(x)}}||{{V}^{*}}-\overline{V}|{{|}_{\infty ,\frac{1}{{{V}_{L}}}}}  \nonumber \\ 
& \qquad \le ||{{V}_{L}}|{{|}_{1,c(x)}}\left( ||{{V}^{*}}-V|{{|}_{\infty ,\frac{1}{{{V}_{L}}}}}\text{+}|\text{ }\!\!|\!\!\text{ }{V-{V}^{*}}|{{|}_{\infty ,\frac{1}{{{V}_{L}}}}} \right) 
  \label{eq55}
\end{align}
where $V_L$ is the Lyapunov function such that $\gamma \mathbb{M} {V}_L<{V}_L$ holds. Based on Lemma \ref{Lemma:6}, 

$$\overline{V}=V-||{{V}^{*}}-V|{{|}_{\infty ,\frac{1}{V_L}}}(\frac{2}{1-{{\Theta }_{V_L}}}-1)$$ is a feasible solution for (\ref{eq30n}), therefore, taking the infinite norm of $V-V^*$ yields
\begin{align}
\begin{array}{l}
{\left\| {{V^{\rm{*}}}{\rm{ - }}V} \right\|_\infty } = {\| {\left\| {{V^{\rm{*}}}{\rm{ - }}V} \right\|{_{\infty \frac{1}{{{V_L}}}}}(\frac{2}{{1 - {\Theta _{{V_L}}}}} - 1){V_L}} \|_\infty } \\
\quad \quad \quad \,\,\,\,\,\,\,\,\,\,\,\, = {\left\| {{V^{\rm{*}}}{\rm{ - }}V} \right\|_\infty }(\frac{2}{{1 - {\Theta _{{V_L}}}}} - 1)
\end{array}
  \label{eqchange}
\end{align}
then, Substituting (\ref{eqchange}) into (\ref{eq55}), (\ref{eq49}) holds.   $\hfill$ $\square$

\vspace{-0.75cm}

\begin{IEEEbiography}[{\includegraphics[width=1in,height=1.25in,clip,keepaspectratio]{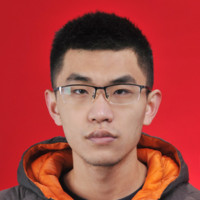}}]{Yuzhen Han} received his B.Sc. degree from Southwest University, China. Since 2019, he has been
working toward Ph.D. degree with the Mechanical Engineering Department, Michigan State University, East Lansing, MI, USA. His  current  research  interests  include  safe control,  reinforcement learning, game theory, and Dempster-Shafer evidence theory.

\end{IEEEbiography}
\vspace{-0.75cm}
\begin{IEEEbiography}[{\includegraphics[width=1in,height=1.25in,clip,keepaspectratio]{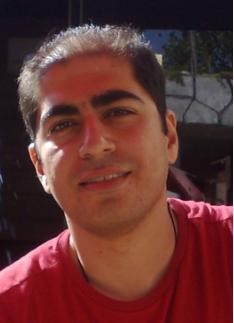}}]{Majid Mazouchi} 
 received his B.Sc. degree in Electrical Engineering from K. N. Toosi University of Technology, Iran, in 2007 and his M.Sc. and Ph.D. degrees in Electrical Engineering from Ferdowsi University of Mashhad, Iran, in 2010 and 2018, respectively. He was a Senior Lecturer at Semnan University, from 2017 to 2018. He is currently a postdoctoral research fellow in Mechanical Engineering Department, Michigan State University, East Lansing, MI, USA. His current research interests include multi-agent systems, reinforcement learning, cyber-physical systems, and distributed control.
\end{IEEEbiography}
\vspace{-0.75cm}
\begin{IEEEbiography}[{\includegraphics[width=1in,height=1.25in,clip,keepaspectratio]{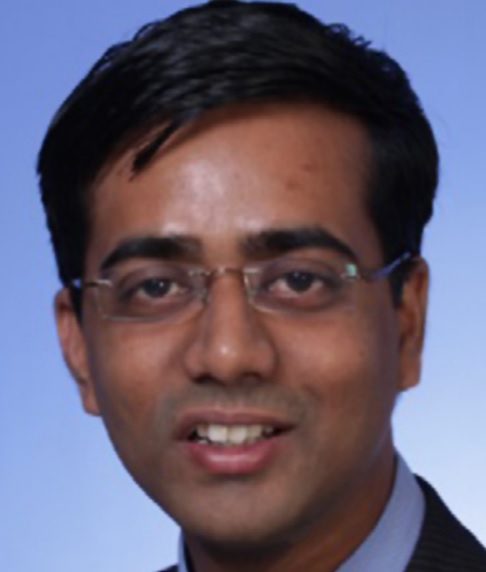}}]{Subramanya P. Nageshrao} 
received the B.E. degree in electronics and communication from Visvesvaraya
Technological University, Belgaum, India, in 2005, M.S. degree in mechatronics from Technische Universität Hamburg-Harburg, Hamburg, Germany, in 2011, and the
Ph.D. degree from Delft University of Technology, Delft, The Netherlands in 2016.
From 2005 to 2009, he was a Software Engineer with Bosch Ltd., Bangalore, India. His current research interests include nonlinear control, distributed control,
machine learning particularly reinforcement learning and its applications for mechatronic and robotic systems.
\end{IEEEbiography}
\vspace{-0.75cm}
\begin{IEEEbiography}[{\includegraphics[width=1in,height=1.25in,clip,keepaspectratio]{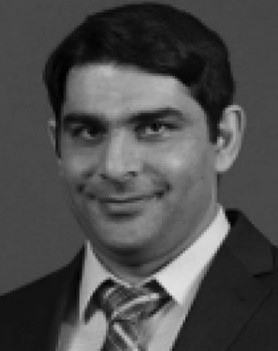}}]{Hamidreza Modares}  (M'15) received the B.S. degree from the University of Tehran, Tehran, Iran, in 2004, the M.S. degree from the Shahrood University of Technology, Shahrood, Iran, in 2006, and the Ph.D. degree from the University of Texas at Arlington, Arlington, TX, USA, in 2015. He was a Senior Lecturer with the Shahrood University of Technology, from 2006 to 2009 and a Faculty Research Associate with the University of Texas at Arlington, Arlington, TX, USA from 2015 to 2016. He is currently an Assistant Professor in Mechanical Engineering Department, Michigan State University, East Lansing, MI, USA. His current research interests include cyber-physical systems, reinforcement learning, distributed control, robotics, and machine learning. Dr. Modares is an Associate Editor for the IEEE TRANSACTIONS ON NEURAL NETWORKS AND LEARNING SYSTEMS. He was the recipient of the best paper award from 2015 IEEE International Symposium on Resilient Control Systems.
\end{IEEEbiography}

\vfill

\end{document}